\documentclass[aip,cha, preprint,showpacs,showkeys]{revtex4-2}
\usepackage{amssymb, dcolumn,  graphicx, latexsym, amsfonts, bm}

\begin{document}
	
\title{Parametric instabilities of the inhomogeneous near SOL tokamak plasma, driven by the coupled  effect of the  high 
harmonic fast wave  and of the ion and electron temperatures gradients, and anomalous heating of  the near SOL ions}
\author{V. V. Mikhailenko}\email[E-mail: ]{vladimir@pusan.ac.kr}
\affiliation{Plasma Research Center, Pusan National University, Busan 46241, South Korea}
\author{V. S. Mikhailenko}\email[E-mail:]{vsmikhailenko@pusan.ac.kr}
\affiliation{Plasma Research Center,  Pusan National University, Busan 46241, South Korea}
\author{Hae June Lee}\email[E-mail: ]{haejune@pusan.ac.kr}
\affiliation{Department of Electrical Engineering, Pusan National University, Busan 46241, South Korea}
	
\date{\today}
\begin{abstract}
Electrostatic parametric instabilities in the inhomogeneous near-SOL tokamak plasma, driven by 
the combined action of a high-harmonic fast wave (HHFW) with a frequency near the 30th ion-cyclotron (IC) harmonic 
and electron and ion temperature gradients, are investigated numerically. The results indicate the parametric decay
of the HHFW into a HHIC (Bernstein) wave and HHIC quasimode. The instability is found 
to exist within a finite wavelength range of the HHFW. The development of the parametric HHIC quasimode decay 
instability leads to the onset of parametric turbulence accompanied by anisotropic ion heating, 
with the ion heating rate across the magnetic field significantly exceeding that along the magnetic field.
\end{abstract}

\maketitle
		
\section{Introduction}\label{sec1}
Radio-frequency (RF) current drive for off-axis current profile control in non-inductively sustained high-confinement (H-mode) 
plasmas is widely regarded as a key requirement for long-pulse and steady-state tokamak operation. 
It has been predicted~\cite{Prater,Li,Pinsker,Wang,Wang1,Taylor} that efficient non-inductive current drive at mid-radius 
can be achieved by the high-harmonic fast wave (HHFW), often referred to as the helicon wave. 
In this regime, the HHFW frequency is $\omega_{0}\sim (30$--$50)\omega_{ci}$, 
i.e. much higher than the ion-cyclotron frequency $\omega_{ci}$ but lower than the lower-hybrid frequency $\omega_{LH}$.

These predictions were based on linear theories of fast-wave heating and helicon current drive (HCD) in the tokamak core plasma. 
They were subsequently supported by low-power ($<0.5$ KW) HCD experiments on the DIII-D~\cite{Pinsker} and Korean 
Superconducting Tokamak Advanced Research (KSTAR)~\cite{Wang,Wang1} tokamaks, as well as by numerical simulations~\cite{Li}. 
However, these studies neglected the effects of HHFW propagation and absorption in the cold, low-density scrape-off layer (SOL) 
and in the narrow pedestal region characterized by strong density and temperature gradients.

These predictions of the linear theory were not confirmed in all FW-heating experiments performed with megawatt levels of RF power. 
Several experiments reported significant losses (50\% or more) of the launched RF power in the plasma edge region, presumably 
in the SOL adjacent to the FW antenna. These losses occurred as the FW propagated away from the antenna. HHFW-heating 
experiments on the National Spherical Torus Experiment (NSTX)~\cite{Biewer,Taylor} confirmed the core-electron heating 
predicted by the linear theory of HHFW propagation and absorption. However, strong edge-ion heating was also 
observed~\cite{Biewer,Hosea} under conditions for which the linear  theory predicted negligible interaction between the HHFW 
and edge ions. It was found\cite{Hosea} that the peak emissivity locations for the heated edge ions are located 
just inside the last closed flux surface (LCFS) in the near-SOL region. Moreover, the edge-ion temperature increased with increasing HHFW power.
The edge-ion heating was strongly anisotropic. The heating rate across the magnetic field substantially exceeded that along the magnetic field. 
These observations could not be explained within the framework of the linear theory.
It was concluded~\cite{Wilson,Biewer, Hosea} that ion-cyclotron (IC) parametric turbulence may provide 
an important nonlinear channel for anomalous FW-power absorption in the SOL. This turbulence 
is driven by the IC-quasimode decay instability observed in NSTX SOL plasma and predicted 
earlier in Refs.~\cite{Porkolab1,Porkolab2} within the dipole approximation. The interaction 
of ions and electrons with the resulting IC parametric turbulence can also explain the observed anisotropic ion heating.

The detailed numerical analysis of parametric instabilities in a uniform SOL plasma driven by HHFW with a 
frequency near the 30th--50th IC harmonic was performed in Ref.~\cite{Mikhailenko1}. 
The results confirmed the experimental observations reported in Refs.~\cite{Taylor,Wilson,Biewer}. 
The numerical analysis demonstrated that the observed instability is consistent with the parametric decay
of the HHFW into a high-harmonic IC (Bernstein) wave and an IC quasimode. 
It was shown~\cite{Mikhailenko1} that the development of this kinetic IC-quasimode decay instability 
is accompanied by anisotropic ion heating. This heating results from the interaction of ions with IC parametric 
turbulence generated by the instability. Under conditions of strong IC damping, the ion-heating 
rate across the magnetic field was found to be much larger than that along the magnetic field. 
In contrast, the anomalous heating of SOL electrons by the IC turbulence was found to be negligibly small.

HHFW heating and current-drive experiments are usually performed in the H-mode confinement regime. 
In this regime, low-frequency drift turbulence, which is responsible for anomalous particle 
and heat transport in the plasma edge, is strongly suppressed inside the last closed flux surface (LCFS).
The turbulence suppression is associated with the development of a poloidal sheared plasma flow. 
As a result, an edge transport barrier is formed. This barrier is characterized by steep density and temperature 
gradients and is commonly referred to as the pedestal. The pedestal separates the cold, low-density SOL plasma from the hot, 
high-density core plasma. Therefore, the results obtained in Ref.~\cite{Mikhailenko1} for a uniform SOL plasma cannot be directly 
applied to the near-SOL pedestal region, where the plasma density and the electron and ion temperatures vary significantly across 
the radial direction.

In this paper, we investigate parametric instabilities and anomalous ion heating 
in an inhomogeneous near-SOL plasma under HHFW heating conditions. The analysis extends the approach 
developed in Refs.~\cite{Mikhailenko1,Mikhailenko2} for a uniform SOL plasma to the more 
realistic case of inhomogeneous density and electron and ion temperature profiles.
Section~\ref{sec2} presents a summary of the theoretical model describing parametric instabilities 
driven by high-power HHFW in the near-SOL plasma. The main result of this analysis, namely the dispersion equation 
for the electrostatic response of the inhomogeneous plasma to the HHFW, is solved numerically in Sec.~\ref{sec3} 
for a reduced three-mode system.
Section~\ref{sec4} develops a renormalized theory of the ion-kinetic-quasimode decay instability. 
This theory includes the nonlinear effect of the  IC resonances broadening caused by ion scattering from the developed 
IC parametric turbulence. The main result of that Section is the level of the electrostatic potential of the IC parametric turbulence, 
powered by the IC quasimode decay instability.  It is then employed in Sec.~\ref{sec5} to investigate the anomalous anisotropic 
ion heating produced by high-harmonic IC parametric turbulence in the near-SOL plasma.
Finally, the conclusions are presented in Sec.~\ref{sec6}.

\section{Theory of parametric instabilities in near-SOL plasma driven by high-power HHFW}\label{sec2}
		
In our previous studies of microscale ion-cyclotron (IC) parametric instabilities driven by fast waves (FW)~\cite{Mikhailenko2} 
and high-harmonic fast waves (HHFW)~\cite{Mikhailenko1}, we adopted the slab-geometry approximation. In this model, the $\hat{x}$, $\hat{y}$, and $\hat{z}$ 
axes correspond to the radial, poloidal, and toroidal directions of the tokamak coordinate system, respectively.
The plasma is assumed to be immersed in a uniform confining magnetic field $\mathbf{B}$ 
directed along the $\hat{z}$ axis and in the electric field of the HHFW, $\mathbf{E}_{0}(\mathbf{\hat{r}},t)$, given by
\begin{eqnarray}
&\displaystyle 
\mathbf{E}_{0}\left(\mathbf{\hat{r}}, t\right) = \mathbf{E}_{0x}\cos\left(\omega_{0} t - \mathbf{k}_{0}\cdot\mathbf{\hat{r}}\right)
+ \mathbf{E}_{0y}\sin\left(\omega_{0} t - \mathbf{k}_{0}\cdot\mathbf{\hat{r}}\right), \label{1}
\end{eqnarray}
where the wavevector $\mathbf{k}_{0} = \mathbf{k}_{0\perp} + \mathbf{k}_{0z}$, 
with $\mathbf{k}_{0\perp} = \mathbf{k}_{0x} + \mathbf{k}_{0y}$ perpendicular to $\mathbf{B}$, and $\mathbf{k}_{0z}$ 
parallel to $\mathbf{B}$.
		
Our theoretical framework is based on the Vlasov equation for the velocity-distribution 
function $F_{\alpha}$ of plasma species $\alpha$, where $\alpha=i$ and $\alpha=e$ denote ions and electrons, respectively,
\begin{eqnarray}
&\displaystyle 
\frac{\partial F_{\alpha}}{\partial t} + \mathbf{\hat{v}} \frac{\partial F_{\alpha}}{\partial \mathbf{\hat{r}}}
+ \frac{e_{\alpha}}{m_{\alpha}}	\left(\mathbf{E}_{0}\left(\mathbf{\hat{r}}, t\right)+ \frac{1}{c}\left[\mathbf{\hat{v}} \times 
\mathbf{B}\right]- \nabla \varphi(\mathbf{\hat{r}}, t)\right) \frac{\partial F_{\alpha}}{\partial \mathbf{\hat{v}}}= 0, 
\label{2}
\end{eqnarray}
together with the Poisson equation for the self-consistent electrostatic potential $\varphi(\mathbf{\hat{r}},t)$ generated by 
the plasma response to the HHFW:
\begin{eqnarray}
&\displaystyle 
\nabla^2 \varphi(\mathbf{\hat{r}}, t) = -4\pi \sum_{\alpha = i, e} e_{\alpha} \delta n_{\alpha}(\mathbf{\hat{r}}, t)
= -4\pi \sum_{\alpha = i, e} e_{\alpha} \int f_{\alpha}(\mathbf{\hat{v}}, \mathbf{\hat{r}}, t) d\mathbf{\hat{v}}. 
\label{3}
\end{eqnarray}
In Eq.~(\ref{3}), $\delta n_{\alpha}(\mathbf{\hat{r}}, t)$ and $f_{\alpha}$ denote the fluctuating parts 
of the density $n_{\alpha}$ and the distribution function $F_{\alpha}$, respectively. Here,
$f_{\alpha}=F_{\alpha}-F_{0\alpha}$, where $F_{0\alpha}$ is the equilibrium distribution function of species $\alpha$. 
Direct application of the spectral transforms with respect to $\mathbf{\hat{r}}$ and $t$ to the linearized 
Vlasov equation~(\ref{2}), together with the Poisson equation~(\ref{3}), leads to an infinite set of coupled equations for the 
wave harmonics generated by the HHFW. These harmonics include the distribution-function components
$f_{\alpha}(\mathbf{\hat{v}},\mathbf{\hat{k}}-n\mathbf{k}_{0},\omega-n\omega_{0})$
with $n=0,\pm1,\pm2,\ldots$, and the electrostatic-potential harmonics
$\varphi(\mathbf{\hat{k}}-m\mathbf{k}*{0},\omega-m\omega_{0})$
with $m=0,\pm1,\pm2,\ldots$.

As it was shown in Ref.~\cite{Mikhailenko2}, the equation for the separate spatial Fourier mode  of the ion Vlasov equation~(\ref{2}) 
can be derived using a moved frame of  references with new position and velocity variables, $\mathbf{r}_i$ and $\mathbf{v}_i$, 
which follow the oscillatory ion trajectory  $\mathbf{\hat{V}}_0(\mathbf{\hat{r}}, t) = \mathbf{V}_{i0}(\mathbf{r}_i, t)$ 
in the magnetic field $\mathbf{B}$ 	and the external HHFW electric field $\mathbf{E}_0(\mathbf{\hat{r}}, t)$. 
In this moved frame, the ion Vlasov equation  contains the external field $\mathbf{E}_0(\mathbf{r}_i, t)$ only in terms proportional 
to the parameter $k_{i0} \xi_i $, where $\xi_i$ is the amplitude of the ion oscillation in $\mathbf{B}$  and  
$\mathbf{E}_0(\mathbf{\hat{r}}, t)$ fields relative to the laboratory frame.  For FW and HHFW fields in typical heating and 
current drive experiments, the ion displacement $\xi_i$ is negligible smaller than the HHFW wavelength $2\pi k_{i0}^{-1}$, 
so these terms can be neglected. Without these small terms, the ion Vlasov equation in $(\mathbf{r}_i, \mathbf{v}_i)$ variables 
reduces to the  familiar form for a magnetized plasma in the absence of external, spatially inhomogeneous, time-dependent 
electromagnetic field~\cite{Mikhailenko2}:

\begin{eqnarray}
\frac{\partial F_i(t, \mathbf{r}_i, \mathbf{v}_i)}{\partial t}
+ \mathbf{v}_i \frac{\partial F_i(t, \mathbf{r}_i, \mathbf{v}_i)}{\partial \mathbf{r}_i}
+ \frac{e_i}{m_i} \left( -\nabla\varphi_i(\mathbf{r}_i, t) + \frac{1}{c}[\mathbf{v}_i \times \mathbf{B}] \right)
\frac{\partial F_i}{\partial \mathbf{v}_{i}} = 0. \label{4}
\end{eqnarray}
The equilibrium ion distribution function $F_{i0}(\mathbf{r}_i,\mathbf{v}_i)$ 
is time-independent and is determined from Eq.~(\ref{4}) by setting $\varphi_i(\mathbf{r}_i,t)=0$.

It follows from Eq.~(\ref{4}) that, in the variables $\mathbf{r}_{i}$ and $\mathbf{v}_{i}$, 
the Vlasov equation for the fluctuating distribution function $f_i=F_i-F_{i0}$ 
has the same form as in a plasma without FW fields. Here, $F_{i0}$ denotes the equilibrium ion distribution function, 
and the electrostatic potential enters through $\varphi_i(\mathbf{r}_i,t)$.
Introducing the guiding-center coordinates $X_i$, $Y_i$, and $z_i$ in the ion oscillating frame by

\begin{eqnarray}
&\displaystyle 
x_{i}=X_{i}-\frac{v_{i\bot}}{\omega_{ci}}\sin \left(\phi_{1}-\omega_{ci}t\right), 
\nonumber 
\\
&\displaystyle
y_{i}=Y_{i}+\frac{v_{i\bot}}{\omega_{ci}}\cos \left(\phi_{1}-\omega_{ci}t\right),
	\label{5}
\end{eqnarray}
where \quad
\begin{eqnarray}
&\displaystyle 
 v_{ix}=v_{i\bot}\cos \phi,  \quad  v_{iy}=v_{i\bot}\sin \phi ,  \quad  \phi=\phi_{1}-\omega_{ci}t, 
\label{6}
\end{eqnarray}
the Vlasov equation for $f_{i}$ has a form
\begin{eqnarray}
&\displaystyle 
\frac{\partial f_{i}}{\partial t}
+\frac{e_{i}}{m_{i}\omega_{ci}}\left(\frac{\partial\varphi_{i}}{\partial X_{i}} \frac{\partial
f_{i}} {\partial Y_{i}}-\frac{\partial\varphi_{i}}{\partial
Y_{i}} \frac{\partial f_{i}} {\partial X_{i}}\right)
\nonumber 
\\
&\displaystyle
+\frac{e_{i}}{m_{i}}\frac{\omega_{ci}}{v_{i\bot}}
\left(\frac{\partial\varphi_{i}}{\partial \phi_{1}} \frac{\partial
f_{i}} {\partial v_{i\bot}}-\frac{\partial\varphi_{i}}{\partial
v_{i\bot}}\frac{\partial f_{i}} {\partial \phi_{1}}\right)-\frac{e_{i}}{m_{i}}
\frac{\partial\varphi_{i}}{\partial z_{i}} \frac{\partial f_{i}}{\partial v_{iz}}
\nonumber 
\\
&\displaystyle
= \frac{e_{i}}{m_{i}\omega_{ci}}\frac{\partial\varphi_{i}}{\partial Y_{i}}\frac{\partial F_{i0}}{\partial X_{i}}
- \frac{e_{i}}{m_{i}}\frac{\omega_{ci}}{v_{i\bot}}\frac{\partial\varphi_{i}}{\partial\phi_{1}}
\frac{\partial F_{i0}}{\partial v_{i\bot}}+\frac{e_{i}}{m_{i}}\frac{\partial\varphi_{i}}{\partial z_{i}}
\frac{\partial F_{i0}}{\partial v_{iz}}. 
\label{7}
\end{eqnarray}
Instead of the linearisation of Eq. (\ref{7}) as a routine initial step to the solution of 
this nonlinear equation, we solve Eq. (\ref{7}) employing the 
procedure of the "renormalized linearisation" which is presented in details in Ref. \cite{Mikhailenko2}. 
This procedure  provides the inclusion to the  derived linear solution for $f_{i}$ the averaged nonlinear effect of the 
scattering of ions by the ensemble of the IC waves,  which is the dominant effect in the saturation 
of the HHIC parametric instabilities\cite{Mikhailenko1, Mikhailenko2}.

The procedure of the "renormalized linearisation" consists in the transformation of the leading center 
coordinates $X_{i}$, $Y_{i}$ and velocity  coordinates $v_{i\bot}$, $\phi_{1}$, $v_{iz}$ in Eq. (\ref{7}) to new 
coordinates $\bar{X}_{i}$, $\bar{Y}_{i}$, and velocity  coordinates $\bar{v}_{i\bot}$, $\bar{\phi}_{1}$, $\bar{v}_{iz}$,
\begin{eqnarray}
&\displaystyle 
X_{i}=\bar{X}_{i}+\delta X_{i}, Y_{i}=\bar{Y}_{i}+\delta Y_{i},v_{i\bot}=\bar{v}_{i\bot}+\delta v_{i\bot}, 
\phi_{1}=\bar{\phi}+\delta\phi, v_{iz}=\bar{v}_{iz}+\delta v_{iz},
\label{8}
\end{eqnarray}
which account for in the explicit form\cite{Mikhailenko2} the distortions $\delta X_{i}$ , 
$\delta Y_{i}$, $\delta v_{i\bot}$, $\delta\phi$, 
$\delta v_{iz}$ of the unperturbed variables $\bar{X}_{i}$, $\bar{Y}_{i}$, $\bar{v}_{i\bot}$, 
$\bar{\phi}$, $\bar{v}_{iz}$ by the electrostatic turbulence,
\begin{eqnarray}
&\displaystyle 
\delta	X_{i}=-\frac{e_{i}}{m_{i}\omega_{ci}}\int\limits^{t}_{t_{0}}\frac{\partial\varphi_{i}}
{\partial \bar{Y}_{i}}dt_{1},
\label{9}
\\
&\displaystyle 
\delta Y_{i}=\frac{e_{i}}{m_{i}\omega_{ci}}\int\limits^{t}_{t_{0}}\frac{\partial\varphi_{i}}
{\partial \bar{X}_{i}}dt_{1},	
\label{10}
\\
&\displaystyle 
\delta v_{i\bot}=
\frac{e_{i}}{m_{i}}\frac{\omega_{ci}}{v_{i\bot}}
\int\limits^{t}_{t_{0}}\frac{\partial\varphi_{i}}{\partial
\bar{\phi}}dt_{1},
\label{11}
\\
&\displaystyle 
\delta\phi=-\frac{e_{i}}{m_{i}}\frac{\omega_{ci}}{v_{i\bot}}
\int\limits^{t}_{t_{0}}\frac{\partial\varphi_{i}}{\partial
\bar{v}_{i\bot}}dt_{1},
\label{12}
\\
&\displaystyle 
\delta v_{iz}	=-\frac{e_{i}}{m}_{i}\int\limits^{t}_{t_{0}}\frac{\partial\varphi_{i}}{\partial
\bar{z}_{i}}dt_{1}.
\label{13}
\end{eqnarray}
Note, that $\bar{X}_{i}$, $\bar{Y}_{i}$, $\bar{v}_{i\bot}$, $\bar{\phi}$, $\bar{v}_{iz}$ 
are the integrals of the system of equations for the characteristics to Eq. (\ref{7}). 

The perturbed electrostatic potential $\varphi_{i}$ is presented in variables $\bar{X}_{i}$, 
$\bar{Y}_{i}$, $\bar{v}_{i\bot}$, $\bar{\phi}$, $\bar{v}_{iz}$ and $\delta X_{i}$,
$\delta Y_{i}$, $\delta v_{i\bot}$, $\delta\phi$, $\delta v_{iz}$ in a form
\begin{eqnarray}
& \displaystyle
\varphi_{i}\left(\mathbf{r}_{i},t \right)=\int 	d\mathbf{k}_{i}d\omega
\varphi_{i}\left(\mathbf{k}_{i},\omega \right)
\nonumber
\\
&\displaystyle 
\times\exp\Big[-i\omega t+ik_{ix}x_{i} +ik_{iy}y_{i}+ik_{iz}z_{i}\Big] 
\nonumber
\\
&\displaystyle 
=\sum_{n=-\infty}^{\infty}\int
d\mathbf{k}_{i}d\omega
\varphi_{i}\left(\mathbf{k}_{i},\omega \right)J_{n}\left(
\frac{k_{i\bot}\bar{v}_{i\bot}}{\omega_{ci}}\right)
\nonumber\\
&\displaystyle 
\times\exp\left(i\mathbf{k}_{i}\delta\mathbf{r}_{i}\left(t\right)\right)
\exp\Big[-i\omega t+ik_{ix}\bar{X}_{i} 	\nonumber\\ & \displaystyle
+ik_{iy}\bar{Y}_{i}+ik_{iz}\left(\bar{z}_{i} 
+\bar{v}_{iz}t\right)-in\left(\bar{\phi}-\omega_{ci}t-\theta \right)\Big],
\label{14}
\end{eqnarray}
where $J_{n}$ is the Bessel function of the order $n$. The nonlinear phase shift $\mathbf{k}_{i}
\delta\mathbf{r}_{i}\left(t\right)$, 
resulted from the perturbations of the ions orbits, 
\begin{eqnarray}
	& \displaystyle \mathbf{k}_{i}\delta\mathbf{r}_{i}\left(t\right)=k_{ix}\delta X_{i}+k_{iy}\delta Y_{i}
	+k_{iz}\int\limits^{t}\delta v_{iz}\left(\tau \right)d\tau 
	\nonumber\\ & \displaystyle
	-\frac{k_{i\bot}\delta v_{i\bot}}{\omega_{ci}}\sin
	\left(\phi-\theta\right)-\frac{k_{i\bot}\bar{v}_{i\bot}}{\omega_{ci}}
	\cos\left(\phi-\theta\right)\delta\phi, 
	\label{15}
\end{eqnarray}
is included in Eq.~(\ref{15}) without terms of second order in $\delta X_{i}$, $\delta Y_{i}$,
$\delta v_{i\perp}$, $\delta v_{iz}$, and $\delta\phi$.
It was shown in Ref.~\cite{Mikhailenko4} that, with the new variables
$t$, $\bar{X}$, $\bar{Y}$, $\bar{z}_{1}$, $\bar{v}_{\bot}$, $\bar{\phi}_{1}$, and $\bar{v}_{z}$,
defined above by Eqs.~(\ref{8})--(\ref{13}), the second-order nonlinearities in the equation for
$f_{i}\left(t_{1}, \bar{X}_{i}, \bar{Y}_{i}, \bar{z}_{1}, \bar{v}_{i\bot}, \bar{\phi}_{1}, \bar{v}_{iz}\right)$
are transformed into third-order nonlinearities with respect to the potential $\varphi$ on the left-hand side of Eq.~(\ref{7}).
Neglecting these nonlinearities, we obtain a linear equation for $f_{i}$ with known
$F_{i0}\left(\bar{v}_{i\bot},\bar{v}_{iz},\bar{X}_{i}\right)$,
\begin{eqnarray}
& \displaystyle 
\frac{\partial f_{i}}{\partial t} =
\frac{e_{i}}{m_{i}}\left[\frac{1}{\omega_{ci}}\frac{\partial\varphi_{i}}{\partial \bar{Y_{i}}}\frac{\partial
F_{i0}}{\partial \bar{X}_{i}} -\frac{\omega_{ci}}{\bar{v}_{i\bot}}
\frac{\partial\varphi_{i}}{\partial \bar{\phi}_{1}} \frac{\partial
F_{i0}}{\partial \bar{v}_{i\bot}} +\frac{\partial\varphi_{i}}{\partial z_{1}}
\frac{\partial F_{i0}}{\partial \bar{v}_{iz}}\right],  
\label{16}
\end{eqnarray}
with the solution 
\begin{eqnarray}
& \displaystyle f_{i}\left(t, \bar{X}, \bar{Y}, \bar{z}_{1}, \bar{v}_{\bot}, \bar{\phi}_{1}, \bar{v}_{z}\right) 
\nonumber
\\
&\displaystyle 
=\frac{e_{i}}{m_{i}}\int\limits^{t}\left[\frac{1}{\omega_{ci}}\frac{\partial\varphi_{i}}{\partial \bar{Y_{i}}}\frac{\partial
	F_{i0}}{\partial \bar{X}_{i}}-\frac{\omega_{ci}}{\bar{v}_{\bot}}
\frac{\partial\varphi_{i}}{\partial \bar{\phi}_{1}} \frac{\partial
F_{i0}}{\partial \bar{v}_{\bot}} +\frac{\partial\varphi_{i}}{\partial z_{1}}
\frac{\partial F_{i0}}{\partial \bar{v}_{z}} \right] dt'. 
\label{17}
\end{eqnarray}
In Eq.~(\ref{17}), we account for the average effect of ion scattering by the electrostatic turbulent field. 
Following the conventional renormalized theory~\cite{Dum}, we employ the simplified approximation 
that particle scattering by IC turbulence is a Gaussian process, for which the relation~\cite{Dum}
\begin{eqnarray}
& \displaystyle
\left\langle e^{i\mathbf{k}_{i\bot}\left(\delta\mathbf{r}\left(t\right)-\delta\mathbf{r}\left(t_{1}\right)\right)}\right\rangle
\simeq e^{-\frac{1}{2}\left\langle\left(\mathbf{k}_{i\bot}\delta\mathbf{r}
\left(t-t_{1}\right)\right)^{2}\right\rangle}=e^{-C_{i}\left(t-t_{1}\right)}
\label{18}
\end{eqnarray}
holds. Equation~(\ref{18}) includes the nonlinear coefficient
$C_{i}$, which is responsible for the random scattering of ions by IC turbulence.
For a Maxwellian ion distribution, this coefficient is given explicitly by Eq.~(\ref{31}).
The solution of Eq.~(\ref{17}) for $f_{i}$, Fourier transformed with respect to $\mathbf{r}_{i}$ 
and accounting for the average effect of ion scattering, is readily obtained and is given by
\begin{eqnarray}
&\displaystyle 
f_{i}\left( v_{i\bot},\phi, v_{iz}, \mathbf{k}_{i}, \bar{X}_{i}, t\right)=\frac{ie_{i}}{m_{i}}\sum\limits_{n=- 	\infty}^{\infty}
\sum\limits_{n_{1}=-\infty}^{\infty} \int\limits_{t_{0}}^{t}dt_{1}\varphi_{i}\left(\mathbf{k}_{i}, t_{1}\right)
\nonumber  
\\
&\displaystyle
\times
\exp\Big[-ik_{iz} \bar{v}_{iz}\left(t-t_{1}\right)-C_{i}\left(t-t_{1}\right)
\nonumber
\\ 
&\displaystyle
+in\left( \bar{\phi}_{1}-\omega_{ci}t-\theta\right)-in_{1}\left(
\bar{\phi}_{1}-\omega_{ci}t_{1}-\theta\right) \Big]
\nonumber
\\ 
&\displaystyle
\times
J_{n}\left(\frac{k_{i\bot}\bar{v}_{i\bot}}{\omega_{ci}}\right) J_{n_{1}}	\left(\frac{k_{i\bot}\bar{v}_{i\bot}}{\omega_{ci}}\right)
\nonumber
\\ 
&\displaystyle
\times
\left[\frac{k_{iy}}{\omega_{ci}}\frac{\partial F_{i0}}{\partial  \bar{X}_{i}} 
+  \frac{n_{1}\omega_{ci}}{\bar{v}_{i\bot}}
\frac{\partial F_{i0}}{\partial \bar{v}_{i\bot}}+ k_{iz}\frac{\partial
F_{i0}}{\partial \bar{v}_{iz}}\right], 
\label{19}
\end{eqnarray}
where $t_{0}\geq 0$ is the initial time.

In this paper, as in Refs.~\cite{Mikhailenko1, Mikhailenko2}, the Poisson equation (\ref{3}) is examined as an equation for 
$\varphi_{i}(\mathbf{k}_{i}, t)$,	obtained by applying the Fourier transform over $\mathbf{r}_{i}$,
\begin{eqnarray}
	& \displaystyle 
	k^{2}_{i}\varphi_{i}\left(\mathbf{k}_{i}, \bar{X}_{i}, t\right)=4\pi\left(e_{i}\delta n_{i}\left(\mathbf{k}_{i}, \bar{X}_{i}, t\right)
	-e\delta n_{e}^{(i)}\left(\mathbf{k}_{i}, \bar{X}_{e},  t\right)\right),
	\label{20}
\end{eqnarray}
where
$\delta n_{i}\left(\mathbf{k}_{i}, \bar{X}_{i}, t\right)=\int d\mathbf{v}_{i}f_{i}\left( v_{i\bot},\phi, v_{iz}, 
\mathbf{k}_{i}, \bar{X}_{i}, t\right)$ is the perturbation of the ion density, Fourier transformed over $\mathbf{r}_{i}$. 
The Fourier transform over $\mathbf{r}_{i}$ should be applied  to the perturbation of the electron density 
$n_{e}(\mathbf{r}_{e}, t)=\int d\mathbf{v}_{e}f_{e}\left( v_{e\bot},\phi, v_{ez}, \mathbf{r}_{e}, \bar{X}_{i}, t\right)$, 
as well as to potential $\varphi_{e}(\mathbf{r}_{e}, t)$, which is involved into the expression for $n_{e}(\mathbf{k}_{e}, t)$, 
using the relations~\cite{Mikhailenko1}: 
\begin{eqnarray}
	& \displaystyle 
	\delta n_{e}^{(i)}\left(\mathbf{k}_{i}, t\right)=\int d\mathbf{r}_{i}\delta n_{e}
	\left(\mathbf{r}_{e}, t\right)e^{-i\mathbf{k}_{i}\mathbf{r}_{i}}
	\nonumber\\
	&\displaystyle 
	=\sum\limits_{m=-\infty}^{\infty}J_{m}\left(a_{ei}\right)e^{im\left(\omega_{0}t+\delta\right)}
	\delta n_{e}^{(e)}\left(\mathbf{k}_{i}-m\mathbf{k}_{0}, t\right),
	\label{21}
\end{eqnarray}
and 
\begin{eqnarray}
	&\displaystyle \varphi_{e}\left(\mathbf{k}_{e},t_{1}\right)=\int d\mathbf{r}_{e}\varphi_{e}
	\left(\mathbf{r}_{e},t_{1}\right)e^{-i\mathbf{k}_{e}\mathbf{r}_{e}}
	\nonumber
	\\ 
	&\displaystyle
	=\sum\limits_{p=-\infty}^{\infty}J_{p}\left(a_{ei}\right) e^{ip\left(\omega_{0}t_{1}+\delta\right)}
	\varphi_{i}\left(\mathbf{k}_{i}-\left(m-p\right)\mathbf{k}_{0}, t_{1}\right).
	\label{22}
\end{eqnarray} 
According to Eq.~(\ref{22}), a single Fourier harmonic $\varphi_{e}(\mathbf{k}_{e}, t)$ of the potential 
$\varphi_{e}(\mathbf{r}_{e}, t)$, as viewed in the ion frame oscillating relative to the electron frame,
appears as an infinite set of harmonics $\varphi_{i}(\mathbf{k}_{i}-(m-p)\mathbf{k}_{0}, t_{1})$.
Two parameters, $a_{ei}$ and $\delta$, are determined by the relations
\begin{eqnarray}
	& \displaystyle 
	a_{ei}=\left\lbrace\left[\sum\limits_{\alpha=i, e}\frac{e_{\alpha}k_{iy}}{2m_{\alpha}\omega_{0}}
	\left(\frac{E_{0x}-E_{0y}}{\omega_{0}-\omega_{c\alpha}} -\frac{E_{0x}+E_{0y}}{\omega_{0}
		+\omega_{c\alpha}}\right)\right]^{2}\right.
	\nonumber
	\\ 
	&\displaystyle
	\left.+\left[\sum\limits_{\alpha=i, e}\frac{e_{\alpha}k_{ix}}{2m_{\alpha}\omega_{0}}
	\left(\frac{E_{0x}+E_{0y}}{\omega_{0}+\omega_{c\alpha}}+\frac{E_{0x}-E_{0y}}{\omega_{0}
		-\omega_{c\alpha}}\right)\right]^{2}\right\rbrace ^{1/2}
	\nonumber
	\\ 
	&\displaystyle
	= |\mathbf{k}_{i}\xi_{ie}|,
	\label{23}
\end{eqnarray}
\begin{eqnarray}
	& \displaystyle 
	\tan \delta=\frac{\sum\limits_{\alpha=i, e}\frac{e_{\alpha}k_{iy}}{m_{\alpha}}\left(\frac{E_{0x}
			+E_{0y}}{\omega_{0}+ \omega_{c\alpha}}+\frac{E_{0x}-E_{0y}}
		{\omega_{0}-\omega_{c\alpha}}\right)}{\sum\limits_{\alpha=i, e}\frac{e_{\alpha}k_{ix}}{m_{\alpha}}
		\left(\frac{E_{0x}+E_{0y}}{\omega_{0}+\omega_{c\alpha}} -\frac{E_{0x}-E_{0y}}
		{\omega_{0}-\omega_{c\alpha}}\right)}.
	\label{24}
\end{eqnarray}
Here, $\xi_{ie}$ is the amplitude of the displacement of electrons relative to ions in HHFW, 
and $\delta$ is the phase of this displacement relative to the wave vector $\mathbf{k}_{i}$ of the microscale 
wave with  $|\mathbf{k}_{i}|\gg |\mathbf{k}_{i0}|$.  
By employing Eqs. (\ref{21}) and (\ref{22}) in the Poisson equation (\ref{20}), the resulted 
functional equation for  $\varphi_{i}\left(\mathbf{k}_{i},t\right)$,  
\begin{eqnarray}
	&\displaystyle 
	\varepsilon\left(\mathbf{k}_{i}, \hat{\omega}\right)\varphi_{i}\left(\mathbf{k}_{i}, 
	\omega \right)+\sum\limits_{q\neq 0}\sum\limits_{m=-\infty}^{\infty}J_{m}\left(a_{ei}\right)
	J_{m+q}\left(a_{ei}\right)e^{iq\delta}
	\nonumber
	\\ 
	&\displaystyle
	\times\varepsilon_{e}\left(\mathbf{k}_{i\bot}, k_{iz}-mk_{0z}, \omega-m\omega_{0}\right)
	\varphi_{i}\left(\mathbf{k}_{i\bot}, k_{iz}+qk_{0z}, \omega+q\omega_{0}\right)=0,
	\label{25}
\end{eqnarray}
where
\begin{eqnarray}
	&\displaystyle 
	\varepsilon\left(\mathbf{k}_{i}, \hat{\omega}\right)=1+\varepsilon_{i}
	\left(\mathbf{k}_{i}, \hat{\omega}\right)+ \sum\limits_{m=-\infty}^{\infty}
	J^{2}_{m}\left(a_{ei}\right)\varepsilon_{e}\left(\mathbf{k}_{i\bot}, k_{iz}-mk_{0z}, 
	\omega-m\omega_{0}\right),
	\label{26}
\end{eqnarray}
was derived in Refs. \cite{Mikhailenko1, Mikhailenko2}. Equations ~(\ref{25}), (\ref{26}), are the basic 
equations of the theory of the parametric instabilities, which include the nonlinear effect of the scattering of ions by the 
parametric turbulence involved in $\varepsilon_{i}\left(\mathbf{k}_{i}, \hat{\omega}\right)$ by the renormalized  frequency 
\begin{eqnarray}
	&\displaystyle 
	\hat{\omega}=\omega-iC_{i}\left(\mathbf{k}_{i\bot}\right) .
	\label{27}
\end{eqnarray}
In Eqs. (\ref{25}), (\ref{26}), only the finite wavelength  $k_{0z}$ of the HHFW along the magnetic field 
is included,  because the unstable perturbations can have a wave number 
component $k_{iz}$ along the magnetic field commensurate with $mk_{0z}$. At the same time, we omit 
$\mathbf{k}_{0\bot}$ in Eqs.~(\ref{25}), (\ref{26}) because the wave vector $\mathbf{k}_{i\bot}$ of the unstable  
perturbations is always larger than $\mathbf{k}_{0\bot}$.

Equation  (\ref{25}) displays, that all Fourier harmonics of the potential $\varphi_{i}$ are linearly coupled due to 
the relative oscillatory motion of plasma species in HHFW. This coupling initiates the development of the IC 
parametric instabilities driven by HHFW. Note, that when the frequency $\omega_{0}$ 
of the strong pumping wave vanishes, Eq. (\ref{25}) transforms to the equation $\varphi_{i}
\left(\mathbf{k}, \omega\right)\left(1+\varepsilon_{i}\left(\mathbf{k},\hat{\omega}\right)+
\varepsilon_{e}\left(\mathbf{k},\omega-\mathbf{k}\mathbf{U}\right)\right)=0$, where 
$\mathbf{U}$ is the stationary velocity of electrons relative to ions. In this case, the coupling the ion mode and 
convected electron mode triggers the development of the linear instabilities driven by a steady current. 
From this point of view, Eq. (\ref{25}) is the general form of the linear equation for electrostatic potential $\varphi_{i}$, 
which governs the electrostatic instabilities of plasma  driven by the relative oscillatory motion of plasma species 
in FW with $|\mathbf{k}_{0}|\ll |\mathbf{k}_{i}|$.	It should be emphasized here that there is a fundamental difference 
between the parametric instabilities driven by the high-power pumping wave considered in this paper, and those 
driven by a low-power wave, where the motion 
of plasma species is negligibly small and plasma perturbations are linearly independent. In the latter case, the 
development of various parametric instabilities is the result of nonlinear mode coupling~\cite{Oraevsky}.

For the Maxwellian distribution $F_{\alpha0}\left(\mathbf{v}_{\alpha}, \bar{X}_{\alpha}\right)$ for ions and electrons 
$\left(\alpha=i, e\right)$,  
\begin{eqnarray}
	&\displaystyle
	F_{\alpha 0}\left(\mathbf{v}_{\alpha}, \bar{X}_{\alpha}\right)=\frac{n_{0\alpha}\left(\bar{X}_{\alpha}\right)}{\left(2\pi 
		v^{2}_{T\alpha}\left(\bar{X}_{\alpha}\right)\right)^{3/2}}\exp \left(-\frac{v^{2}_{i}}{2v^{2}_{T\alpha}
		\left(\bar{X}_{\alpha}\right)}\right),
	\label{28}
\end{eqnarray}
with inhomogeneous density $n_{0\alpha}\left(\bar{X}_{\alpha}\right)$ and  temperatures $T_{\alpha}\left(\bar{X}_{\alpha}\right)$,
the renormalized ion dielectric permittivity $\varepsilon_{i}\left(\mathbf{k}_{i}, \hat{\omega}\right)$, which  
includes  the averaged nonlinear effect of ions  scattering by the ensemble of the IC waves\cite{Mikhailenko2}, 
and the linear electron dielectric permittivity $\varepsilon_{e}\left(\mathbf{k}_{i}, \omega\right)$ are
\begin{eqnarray}
	&\displaystyle 
	\varepsilon_{i}\left(\mathbf{k}_{i}, \hat{\omega}\right)=
	\frac{1}{k_{i}^2\lambda _{Di}^2}\left[1+ i\sqrt \pi \left( z_{i0}-\xi_{i}\left(1-\frac{\eta_{i}}{2} \right) \right) 
	\sum\limits_{p = - \infty }^{\infty}W\left(z_{ip}\right)A_{ip}\left(k^{2}_{i\bot}\rho^{2}_{i}\right)\right.
	\nonumber
	\\ 
	&\displaystyle
	-\eta_{i}\xi_{i}\sum\limits_{p = - \infty }^{\infty}
	z_{ip}\left(1+i\sqrt{\pi}z_{ip} W\left(z_{ip}\right)\right) A_{ip}\left(k^{2}_{i\bot}\rho^{2}_{i}\right)
	\nonumber
	\\ 
	&\displaystyle
	+\left. \eta_{i}\xi_{i}\sum\limits_{p = - \infty }^{\infty}
	i\sqrt{\pi} W\left(z_{ip}\right)k^{2}_{i\bot}\rho^{2}_{i}
	\left(  A_{ip}\left(k^{2}_{i\bot}\rho^{2}_{i}\right)-\hat{A}_{ip}\left(k^{2}_{i\bot}\rho^{2}_{i}\right)\right) 
	\right],
	\label{29}
\end{eqnarray}

\begin{eqnarray}
	&\displaystyle 
	\left. \varepsilon_{e}\left(\mathbf{k}_{i}, \omega-m\omega_{0}\right)
	=\frac{1}{k_{i}^2\lambda_{De}^2}
	\right\{ 1+	\xi_{e}\eta_{e}z_{em}A_{e0}\left(k^{2}_{i\bot}\rho^{2}_{e}\right)
	\nonumber
	\\ 
	&\displaystyle
	+\left. i\sqrt{\pi}z_{em}W\left(z_{em}\right)
	\left[ \left(z_{em}-\xi_{e}\left( 1-\frac{\eta_{e}}{2}\right)-\xi_{e}\eta_{e}z_{em}^{2}  \right)
	A_{e0}\left(k^{2}_{i\bot}\rho^{2}_{e}\right)
	\right. \right. 
	\nonumber
	\\ 
	&\displaystyle
	\left. 
	+\xi_{e}\eta_{e}k^{2}_{i\bot}\rho^{2}_{e}
	\left(A_{e0}\left(k^{2}_{i\bot}\rho^{2}_{e}\right)-A_{e1}\left(k^{2}_{i\bot}\rho^{2}_{e}\right)\right) 
	\Big] 
	\right\} .
	\label{30}
\end{eqnarray}
In Eqs. (\ref{29}), (\ref{30}), $\lambda_{Di(e)}$ is the ion (electron) Debye length, 
$v^{2}_{T\alpha}={T_{\alpha}}/m_{\alpha}$,  $\rho_{i}= v_{Ti}/\omega_{ci}$ 
is the ion thermal Larmor radius, $I_{p}$ is the modified Bessel function of order $p$, 
\begin{eqnarray}
	&\displaystyle 
	A_{\alpha 
		p}\left(k^{2}_{i\bot}\rho^{2}_{\alpha}\right)=I_{p}\left(k^{2}_{i\bot}\rho^{2}_{\alpha}\right)
	e^{-k^{2}_{i\bot}\rho^{2}_{\alpha}},\nonumber
	\\ 
	&\displaystyle\hat{A}_{ip}\left(k^{2}_{i\bot}\rho^{2}_{i}\right)=\frac{1}{2}
	\left[I_{p-1}\left(k^{2}_{i\bot}\rho^{2}_{i}\right)+I_{p+1}\left(k^{2}_{i\bot}\rho^{2}_{i}\right)\right]
	e^{-k^{2}_{i\bot}\rho^{2}_{i}}, 
	\nonumber
\end{eqnarray}
$\xi_{\alpha}=k_{y}v_{d\alpha}/\sqrt{2}k_{z}v_{T\alpha}$, 
$\eta_{\alpha}=d \ln T_{\alpha}/d \ln n_{\alpha}$,
$v_{d\alpha}=\left( eT_{\alpha}/e_{\alpha}B_{0}n_{0}\right) \: \left( dn_{0}/d\bar{X}_{\alpha}\right)$, $\alpha=i,e$. \\
$W(z)=e^{-z^{2}}\left(1 +\frac{2i}{\sqrt{\pi}}\int_{0}^{z} e^{t^{2}}dt\right)$ is the complex error function, 
$z_{e} = \frac{\omega}{\sqrt{2}k_{iz}v_{Te}}$, $\zeta_{e} = \frac{\omega_{0}}{\sqrt{2}k_{iz}v_{Te}}$, 
$z_{em}=z_{e}-m\zeta_{e}$, $z_{ip} =\frac{\hat{\omega}-p\omega_{ci}}{\sqrt{2}k_{iz}v_{Ti}}$.  The  nonlinear term 
$C_{i}\left(\mathbf{k}_{i\bot}\right)$, which is responsible  for the effect of the random scattering 
of ions by IC turbulence, for the Maxwellian distribution of ions  is  determined by the equation~\cite{Mikhailenko2}
\begin{eqnarray}
	& \displaystyle C_{i}\left(\mathbf{k}_{i}\right)=\frac{e_{i}^{2}}{m_{i}^{2}\omega^{2}_{ci}}
	Re \sum\limits_{p_{1}=-\infty}^{\infty}\int d\mathbf{k}_{i1}|\varphi_{i}\left(\mathbf{k}_{i1}\right)|^{2}
	\mathcal{F}_{ip_{1}}\left( k_{i\bot}, k_{i1\bot}\right)
	\nonumber
	\\
	&\displaystyle 
	\times
	\sqrt{\frac{\pi}{2}}\frac{1}{k_{i1z}v_{Ti}}
	W\left(\frac{\hat{\omega}-p_{1}\omega_{ci}}{\sqrt{2}k_{i1z}v_{Ti}}\right),
	\label{31}
\end{eqnarray}
where $\mathcal{F}_{ip_{1}}\left( k_{i\bot}, k_{i1\bot}\right)$ is a numerical factor equal to
\begin{eqnarray}
	& \displaystyle \mathcal{F}_{ip_{1}}\left( k_{i\bot}, k_{i1\bot}\right)
	=e^{-k_{i1\bot}^{2}\rho^{2}_{i}}\left[\left(k_{ix}k_{i1y}-k_{iy}k_{i1x}\right)^{2}I_{p_{1}}
	\left(k^{2}_{i1\bot}\rho^{2}_{i}\right)\right.
	\nonumber\\ 
	& \displaystyle
	\left.+\frac{1}{4}k^{2}_{i\bot}k^{2}_{i1\bot}\Big(I_{p_{1}+1}\left(k^{2}_{i1\bot}
	\rho^{2}_{i}\right)+I_{p_{1}-1}\left(k^{2}_{i1\bot}\rho^{2}_{i}\right)\Big)\right].
	\label{32}
\end{eqnarray}
Equation~(\ref{25}) is  an infinite system of equations for the potential
$\varphi_{i}\left(\mathbf{k}_{i}, \omega\right)$ and for the infinite number of harmonics 
$\varphi_{i}\left(\mathbf{k}_{i\bot}, k_{iz}+qk_{0z}, \omega+q\omega_{0}\right)$. By replacing  
$\omega$ with $\omega+q\omega_{0}$ and $\mathbf{k}_{i}$ with $\mathbf{k}_{i}+q\mathbf{k}_{0z}$ in Eq.~(\ref{25}), 
where $q$ is an integer, Eq. (\ref{25}) can be presented in the form of the infinite system
\begin{eqnarray}
	\sum\limits_{q=-\infty}^{\infty}b_{mq}\varphi_{i}\left(\mathbf{k}_{i}-q\mathbf{k}_{0z},  \omega-q\omega_{0}\right)=0,
	\label{33}
\end{eqnarray}
where $m$ and $q$ are integers and the coefficients $b_{mq}$ are determined by the relation
\begin{eqnarray}
	&\displaystyle
	b_{mq}=\delta_{mq}+\left(1+\varepsilon_{i}\left(\mathbf{k}_{i}-m\mathbf{k}_{0z}, 
	\hat{\omega}-m\omega_{0}\right)\right)^{-1}
	\nonumber
	\\ &
	\displaystyle
	\times\sum\limits_{r=-\infty}^{\infty}e^{i\left(m-q\right)\left(\pi+\delta\right)}
	J_{r+m}\left(a\right)J_{r+q}\left(a\right)
	\nonumber
	\\ &
	\displaystyle
	\times 	\varepsilon_{e}\left(\mathbf{k}_{i}+r\mathbf{k}_{0z}, \omega+r\omega_{0}\right).
	\label{34}
\end{eqnarray}	
Because the ion dielectric permittivity  (\ref{29}) with frequency $\hat{\omega}$ has a form as for the linear ion dielectric 
permittivity, the changing of  $\omega$ on $\hat{\omega}$ in the electron dielectric permittivity (\ref{30})  introduces 
negligibly small error in  Eq. (\ref{33})  for  $b_{mq}$ because 
$|z_{e}|=|\omega\left(\mathbf{k}_{i}\right)/\sqrt{2}k_{z}v_{Te}\ll 1$ for the detected instability and $|C_{i}|\ll |\omega|$. 
The equality to zero of the determinant of this homogeneous system,
\begin{eqnarray}
	&\displaystyle
	\text{det}\left\|b_{mq}\left(\mathbf{k}_{i},  \hat{\omega}\right)\right\| =0,
	\label{35}
\end{eqnarray}
gives the linear general dispersion equation for the system (\ref{33}) for the renormalized frequency $\hat{\omega}$, 
the solution  $\hat{\omega}=\Omega\left(\mathbf{k}_{i}\right)$ of which determines the dispersive properties of the 
parametric instabilities driven by the  applied powerful FW.  

The derivation of the complete set of solutions to the dispersion equation~(\ref{35}) 
for the frequency $\Omega\left(\mathbf{k}_{i}\right)$ is a formidable problem  
for the numerical solution of Eq.~(\ref{35}), as it involves the coupling of an infinite 
number of IC waves with frequencies $\omega\left(\mathbf{k}_{i}\right)=n\omega_{ci}+\delta\omega 
\left(\mathbf{k}_{i}\right)$ to an infinite number of waves with wave numbers $\mathbf{k}
_{i}-m\mathbf{k}_{0z}$, $\mathbf{k}_{i}+r\mathbf{k}_{0z}$ and frequencies 
$\omega\left(\mathbf{k}_{i}\right)-m\omega_{0}$ and  $\omega\left(\mathbf{k}_{i}\right)+r\omega_{0}$. 
In this paper, we present results of the numerical solution to Eq.~(\ref{35}) for the 
reduced three-mode system~(\ref{33}), which includes HHIC mode $\varphi_{i}\left(\mathbf{k}_{i}, 
\omega\left(\mathbf{k}_{i}\right)\right)$  ($q=0$), and its satellites $\varphi_{i}\left(\mathbf{k}_{i}-\mathbf{k}_{0z}, 
\omega\left(\mathbf{k}_{i}\right)-\omega_{0}\right)$ ($q=1$) and $\varphi_{i}
\left(\mathbf{k}_{i}+\mathbf{k}_{0z}, \omega\left(\mathbf{k}_{i}\right)+\omega_{0}\right)$ 
($q=-1$). The numerical solution to the reduced  Eq.~(\ref{33}),  was derived, as well as in Ref. \cite{Mikhailenko1}, 
by employing the complex Halley method~\cite{Yau}, which is an extension of the 
Newton method that incorporates the second derivative in the Newton recurrence relation~\cite{Traub} to accelerate 
the procedure of finding complex roots of equations with complex coefficients. In this paper, we focus our attention 
on the not explored yet near SOL region of the pedestal which is 
distinguished by its relatively high plasma density gradient, combined with  gradients of both ion and electron temperatures. 
The interplay of these parameters establishes the near SOL region as a particularly favorable 
environment for the excitation of high-harmonic ion cyclotron (IC) parametric instabilities driven by the coupled effect 
of HHFW  and gradient of the plasma density and of the ion and electron temperature gradients.

\section{Numerical analysis of HHFW-driven parametric instabilities in near-SOL plasma}\label{sec3}

In this section, we present the results of the numerical solution of the dispersion equation~(\ref{35}) for near-SOL plasma driven by a high-power HHFW. 
The HHFW frequency is taken as $\omega_{0}=30.5\,\omega_{ci}$, which lies within the range $\omega_{0}\sim(30$--$50)\omega_{ci}$ 
considered in helicon current-drive studies.
Our numerical investigation of the reduced system~(\ref{35}) was first performed under the simplifying assumption 
of a spatially uniform HHFW ($k_{0z}=0$). Under this assumption, the IC parametric instability 
driven by inverse electron Landau damping, previously identified in Ref.~\cite{Mikhailenko1}, 
is suppressed throughout the HHFW frequency range $\omega_{0}\sim(30$--$50)\omega_{ci}$.
This result indicates that, in order to obtain physically meaningful solutions and capture the relevant physics, 
it is essential to account for the finite wavelength of the HHFW, i.e., to consider the case $k_{0z}\neq0$. 
For the numerical analysis presented below, we adopt representative plasma parameters characteristic of the SOL 
and near-SOL region. The ion and electron densities are taken as $n_{0}=2\times10^{11}$~cm$^{-3}$, 
and the ion-to-electron temperature ratio is set to $T_{i}/T_{e}=1$. 
The mass ratio is chosen as $m_{i}/m_{e}=2\times1840$, corresponding to deuterium ions.
The equilibrium magnetic field is assumed to be $B_{0}=1$~T.
The dimensionless parameters characterizing the parallel wavelengths of the pump and unstable modes 
are chosen as $(k_{0z}\rho_{i})^{-1}=16.763$ and $(k_{z}\rho_{i})^{-1}=14.92$, respectively. 

We use the normalized HHFW electric-field amplitudes $\hat{E}_{0x}$ and $\hat{E}_{0y}$ from Ref.~\cite{Mikhailenko1},
\begin{eqnarray}
&\displaystyle 
\hat{E}_{0x,0y}=\frac{E_{0x,0y}}{\sqrt{4\pi n_{0i}T_{i}}}, 
\nonumber
\end{eqnarray} 
where all parameters are expressed in CGS units, or in the form 
\begin{eqnarray}
&\displaystyle 
\hat{E}_{0x,0y}\approx 750\frac{E_{0x, 0y}}{\sqrt{n_{0i}T_{i}}},
\label{36}
\end{eqnarray} 
when $E_{0x}$ and $E_{0y}$ are measured in V/cm, the ion density is measured in cm$^{-3}$, 
and the ion temperature $T_{i}$ is measured in electron volts.
We use $\hat{E}_{0x}=0.22$, which corresponds to $E_{0x}=400$~V/cm for the values of $n_{0i}$ and $T_{i}$ specified above.
Throughout the numerical simulations, we systematically include all terms in the range $-50\leqslant n\leqslant 50$ 
in the summation over $n$ in Eq.~(\ref{34}) for $\varepsilon_{i}$, as well as all terms in the 
range $-50\leqslant r\leqslant 50$ in the summation over $r$ in Eq.~(\ref{35}) for $b_{mq}$.
This comprehensive approach ensures that the key features of the parametric-instability spectrum 
are accurately represented and that the numerical results are robust with respect to the inclusion of higher-order harmonic effects.

\begin{figure}[!htbp]
\includegraphics[width=0.45\textwidth]{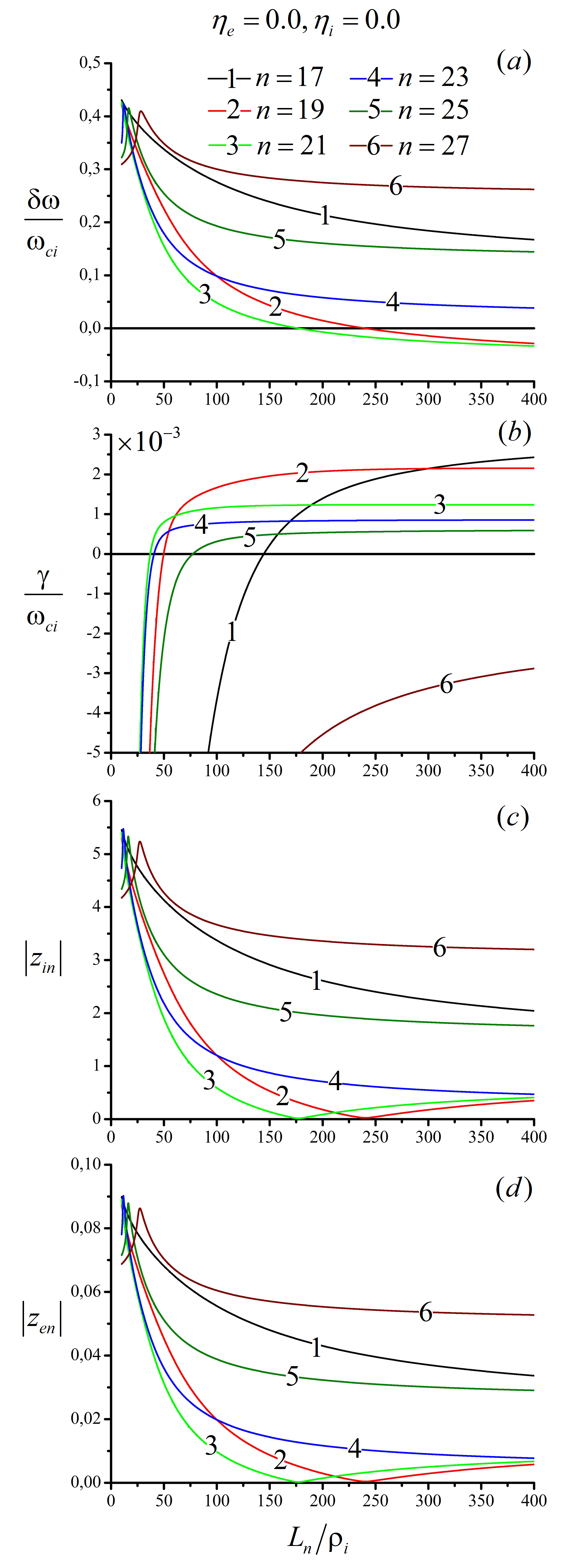}
\caption{\label{fig1} IC harmonic characteristics versus $L_n/\rho_i$ in the near--SOL region for $\eta_e = 0.0$ and $\eta_i = 0.0$.}
\end{figure}

\begin{figure}[!htbp]
\includegraphics[width=0.45\textwidth]{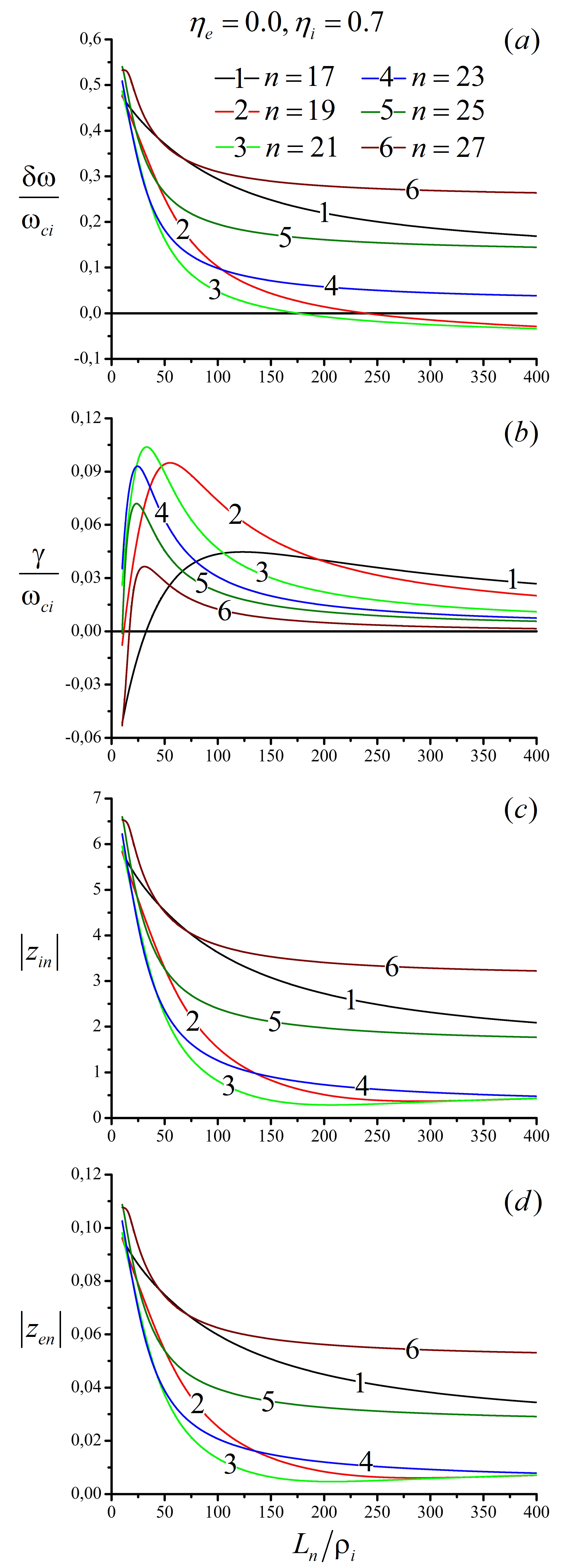}
\caption{\label{fig2} IC harmonic characteristics versus $L_n/\rho_i$ in the near--SOL region for $\eta_e = 0.0$ and $\eta_i = 0.7$.}
\end{figure}

\begin{figure}[!htbp]
\includegraphics[width=0.5\textwidth]{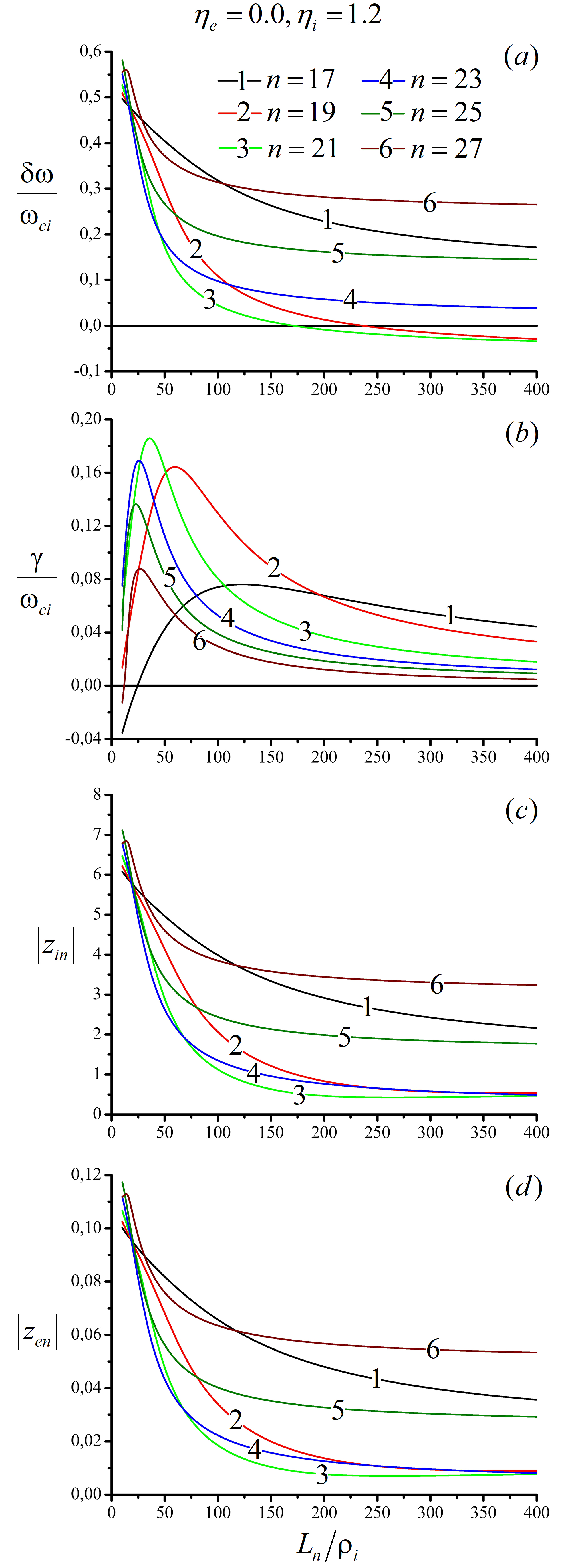}
\caption{\label{fig3} IC harmonic characteristics versus $L_n/\rho_i$ in the near--SOL region for $\eta_e = 0.0$ and $\eta_i = 1.2$.}
\end{figure}

\begin{figure}[!htbp]
\includegraphics[width=0.5\textwidth]{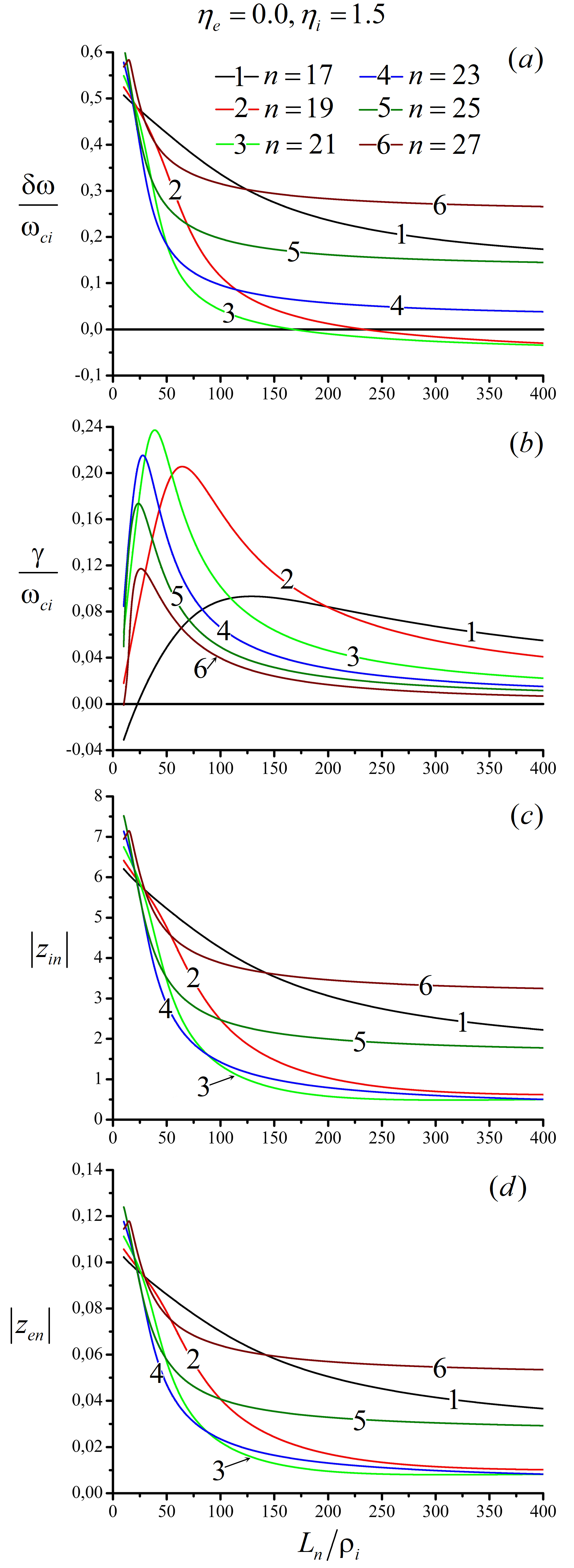} %sol_n27
\caption{\label{fig4} IC harmonic characteristics versus $L_n/\rho_i$ in the near--SOL region for $\eta_e = 0.0$ and $\eta_i = 1.5$.}
\end{figure}

\begin{figure}[!htbp]
	\includegraphics[width=0.5\textwidth]{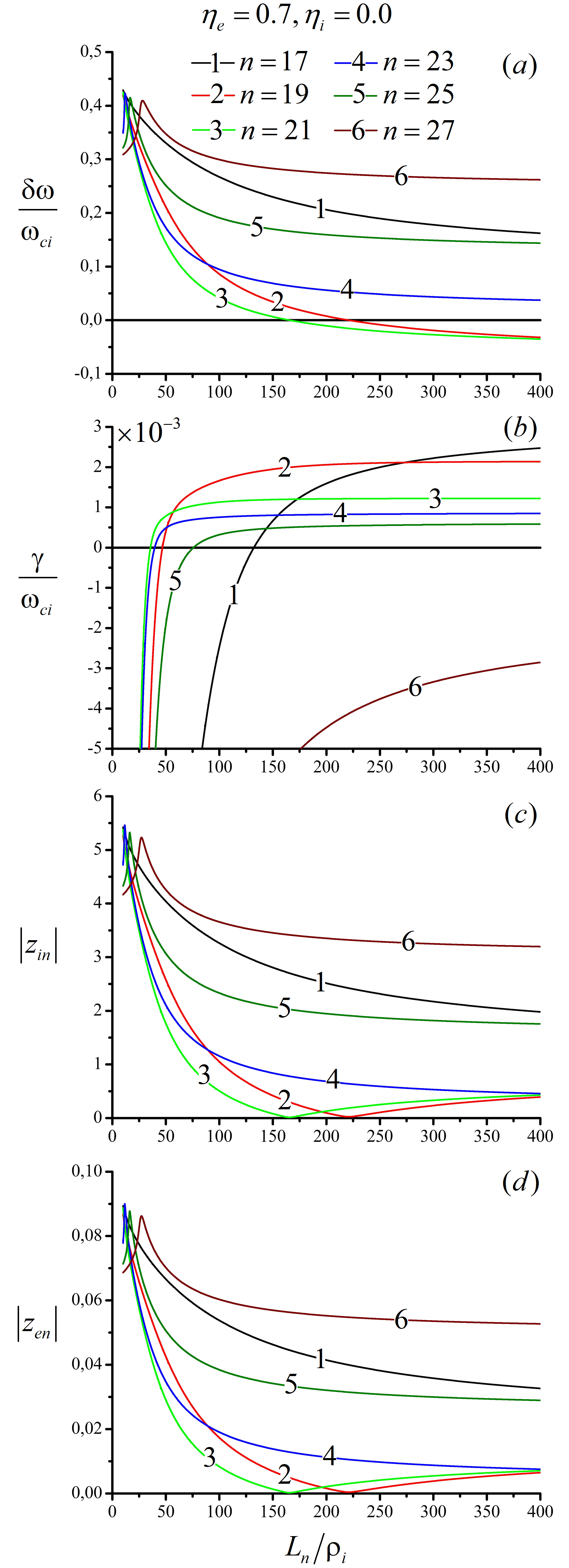} %sol_n27
	\caption{\label{fig5} IC harmonic characteristics versus $L_n/\rho_i$ in the near--SOL region for $\eta_e = 0.7$ and $\eta_i = 0.0$.}
\end{figure}

\begin{figure}[!htbp]
	\includegraphics[width=0.5\textwidth]{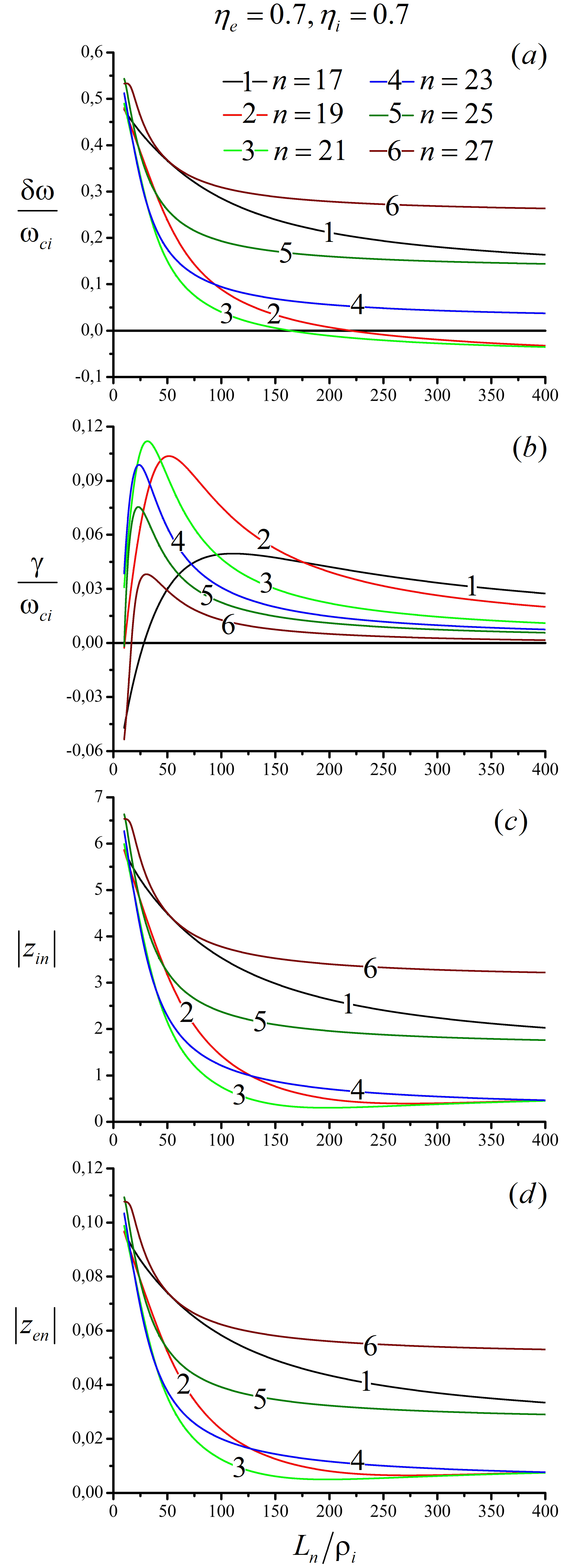} %sol_n27
	\caption{\label{fig6} IC harmonic characteristics versus $L_n/\rho_i$ in the near--SOL region for $\eta_e = 0.7$ and $\eta_i = 0.7$.}
\end{figure}

\begin{figure}[!htbp]
	\includegraphics[width=0.5\textwidth]{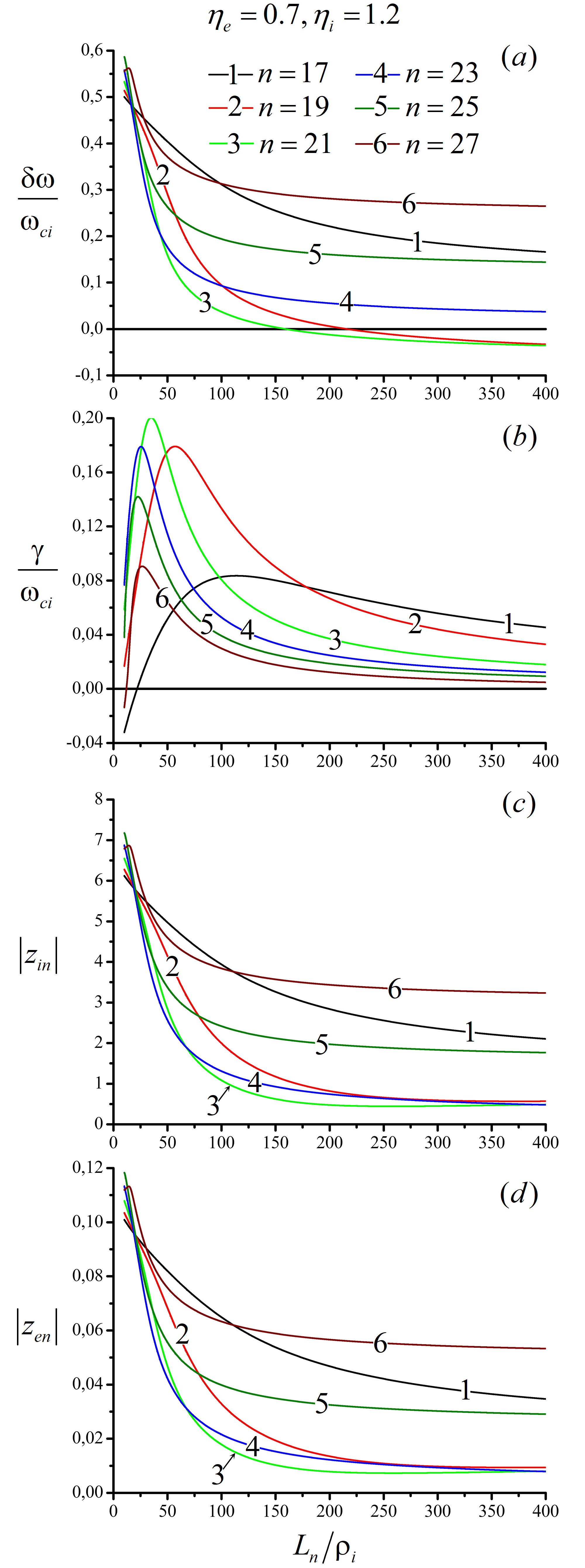} %sol_n27
	\caption{\label{fig7} IC harmonic characteristics versus $L_n/\rho_i$ in the near--SOL region for $\eta_e = 0.7$ and $\eta_i = 1.2$.}
\end{figure}

Figures~\ref{fig1}--\ref{fig7} present the results of the numerical analysis of the reduced system~(\ref{33}) for near-SOL plasma. 
Compared with our previous paper~\cite{Mikhailenko1}, the present study additionally includes the IC harmonic with $n=17$, 
extending the range of high-harmonic ion-cyclotron (IC) modes under consideration. 
The set of harmonics investigated consists of $n=17,\,19,\,21,\,23,\,25,$ and $27$, allowing a direct comparison with the 
previously studied uniform-SOL case. In all calculations, we used $k_{x}\rho_{i}=0.612$ and $k_{y}\rho_{i}=4.65$ values for which 
the IC parametric instability driven by HHFW in the uniform SOL plasma  attains the maximum growth rate.  This baseline case serves as a 
reference point for the subsequent exploration of the HHIC parametric instability in  near-SOL plasma, in which spatial 
inhomogeneities play a decisive role.

Each figure consists of four panels. Panel~(a) shows the normalized frequency deviation, 
$\delta\omega/\omega_{ci}=\mathrm{Re}\,\omega\left(\mathbf{k}_{i}\right)/\omega_{ci}-n$, of the wave frequency 
$\mathrm{Re}\,\omega\left(\mathbf{k}_{i}\right)$ from the $n$-th ion cyclotron harmonic.
Panel~(b) shows the normalized growth rate,
$\gamma/\omega_{ci}=\mathrm{Im}\,\omega\left(\mathbf{k}_{i}\right)/\omega_{ci}$.
Panels~(c) and~(d) display the moduli $|z_{in}|$ and $|z_{en}|$ of the arguments of the plasma-dispersion functions
$W\left(z_{in}\right)$ and $W\left(z_{en}\right)$, respectively.
In all figures, curves~(1)--(6) correspond to the IC harmonics $n=17,\,19,\,21,\,23,\,25,$ and $27$, respectively.

Figure~\ref{fig1} corresponds to parametric drift instabilities in near-SOL plasma with an inhomogeneous density profile and uniform ion and 
electron temperatures, $\eta_e=0.0$ and $\eta_i=0.0$.
Figures~\ref{fig2}--\ref{fig4} illustrate parametric ITG instabilities for $\eta_i=0.7$, $1.2$, and $1.5$, respectively.
Figure~\ref{fig5} presents parametric ETG instabilities for $\eta_e=0.7$.
Figures~\ref{fig6} and~\ref{fig7} show parametric instabilities driven by the combined action of ITG and ETG effects for $\eta_e=0.7$ with $\eta_i=0.7$ and $1.2$, respectively.
Figures~\ref{fig2}--\ref{fig4} and~\ref{fig6}--\ref{fig7}, corresponding to cases with finite ion-temperature gradients, reveal the following features:

\begin{enumerate}
	\item The growth rates of all considered IC harmonics exhibit pronounced maxima at relatively small values of $L_{n}/\rho_{i}$, 
	indicating that the instability is most efficiently excited in plasmas with steep density gradients.
	
	\item The dependence of the maximum growth rate on the harmonic number is non-monotonic. 
	The largest growth rates are obtained for the intermediate harmonics $n=19$--$25$, 
	whereas the edge harmonics $n=17$ and $n=27$ exhibit growth rates approximately three times smaller.
	
	\item Unstable solutions exist only within a restricted range of harmonic numbers. For $\omega_{0}=30.5\,\omega_{ci}$, 
	unstable solutions are found only for the high-harmonic IC modes with $n=17$--$27$, whereas all other harmonics remain stable. 
	The observed decrease of the growth rates toward both ends of the unstable harmonic interval explains the existence of this limited range of unstable harmonics.
	
	\item The instabilities develop in the regime $|z_{in}|<3.5$, where the ion Landau resonance remains strong. 
	This indicates that the considered modes are governed by inverse ion Landau damping. 
	At the same time, $|z_{en}|\ll1$ in all cases, indicating that the electron kinetic contribution remains weak.
\end{enumerate}

All figures demonstrate that the contribution of ETG effects is considerably weaker than that of ITG effects. 
Even for the relatively weak ion-temperature gradient, $\eta_i=0.7$, the growth rates are significantly higher 
than those obtained for $\eta_i=0.0$ (Figures~\ref{fig1} and~\ref{fig5}). Furthermore, for $\eta_i=1.5$ (Figure~\ref{fig4}), 
the maximum growth rates of the intermediate harmonics $n=19$--$25$ are approximately twice 
those obtained for $\eta_i=0.7$ (Figure~\ref{fig2}). These results indicate that ITG effects provide the dominant contribution to the excitation of the IC harmonics investigated in the present study.

A comparison of Figures~\ref{fig1} and~\ref{fig5} shows that the electron-temperature gradient
has only a weak influence on the considered IC harmonics. The corresponding growth rates remain 
at the level of $\gamma/\omega_{ci}\sim10^{-3}$, confirming that ETG effects alone do not provide a significant drive for these modes.

\section{Renormalized theory of the HHIC quasimode decay instability}\label{sec4}

The solutions for the frequency $\hat{\omega}\left(\mathbf{k}_{i}\right)$ presented above are not solutions of the equation
$\varepsilon\left(\mathbf{k}_{i},\hat{\omega}\right)=0$,
where $\varepsilon\left(\mathbf{k}_{i},\hat{\omega}\right)$ is determined by Eq.~(\ref{25}).
In this case, the parametric HHIC instability investigated here may correspond to the ion-kinetic IC-quasimode decay instability. 
Its linear theory was developed  first in Refs.~\cite{Porkolab3,Porkolab4} and subsequently extended to arbitrary values of the parameter 
$a_{ei}$ in Ref.~\cite{Mikhailenko2}.
In the theory of the quasimode decay instability, it is assumed~\cite{Porkolab3,Porkolab4} that, for some $q=q_{0}$, the relation
\begin{eqnarray}
&\displaystyle 
\varepsilon\left(\mathbf{k}_{i}+q_{0}\mathbf{k}_{0z}, \hat{\omega}+q_{0}\omega_{0}\right)=0
\label{37}
\end{eqnarray}
holds in Eq. (\ref{25}) for potential $\varphi_{i}\left(\mathbf{k}_{i}+q_{0}\mathbf{k}_{0z}, 
\hat{\omega}+q_{0}\omega_{0}\right)$ with solution
\begin{eqnarray}
&\displaystyle 
\hat{\omega}=\Omega\left(\mathbf{k}_{i}+q_{0}\mathbf{k}_{0z}\right)-q_{0}\omega_{0}.
\label{38}
\end{eqnarray}
This solution was considered in Ref.~\cite{Mikhailenko2} as the zeroth-order approximation to the solution of Eq.~(\ref{25}).

The next-order approximation, valid for
$\gamma\left(\mathbf{k}_{i}\right)<|\delta\omega\left(\mathbf{k}_{i}\right)|$,
is obtained from Eq.~(\ref{25}) for the quasimode potential
$\varphi_{i}\left(\mathbf{k}_{i}+q_{0}\mathbf{k}_{0z},\hat{\omega}+q_{0}\omega_{0}\right)$,
retaining only the term with $q=q_{0}$ in the summation over $q$.
This approximation yields the growth rate
$\gamma\left(\mathbf{k}_{i}+q_{0}\mathbf{k}_{0z}\right)$
of the ion-kinetic quasimode decay instability~\cite{Mikhailenko2},
\begin{eqnarray}
&\displaystyle \gamma\left(\mathbf{k}_{i}+q_{0}\mathbf{k}_{0z}\right) \approx \left(\frac{\partial \text{Re}\, \varepsilon
\left(\mathbf{k}_{i}+q_{0}\mathbf{k}_{0z}, \Omega\left(\mathbf{k}_{i}+q_{0}\mathbf{k}_{0z}\right)\right)}{\partial \Omega\left(\mathbf{k}_{i}	+q_{0}\mathbf{k}_{0z}\right)}\right)^{-1}
\nonumber 
\\
&\displaystyle
\times
\Big[ -\text{Im}\,\varepsilon
\left(\mathbf{k}_{i}+q_{0}\mathbf{k}_{0z}, \Omega\left(\mathbf{k}_{i}+q_{0}\mathbf{k}_{0z}\right)\right)
\nonumber
\\ 
&\displaystyle
+\sum\limits_{m=-\infty}^{\infty}\sum\limits_{m_{1}=-\infty}^{\infty}J_{m}\left(a_{ei}\right)J_{m+q_{0}}\left(a_{ei}\right) 
J_{m_{1}}\left(a_{ei}\right)J_{m_{1}-q_{0}}\left(a_{ei}\right)
\label{39}
\\
&\displaystyle
\left.
\times
\text{Im}\,\left(\frac{\varepsilon_{e}\left(\mathbf{k}_{i}- m\mathbf{k}_{0z}, 
\Omega\left(\mathbf{k}_{i}+q_{0}\mathbf{k}_{0z}\right)-\left(m+q_{0}\right)\omega_{0}\right)}
{\varepsilon\left(\mathbf{k}_{i}, \Omega\left(\mathbf{k}_{i}+
q_{0}\mathbf{k}_{0z}\right)-q_{0}\omega_{0}\right)}	\right.\right.
\nonumber
\\ 
&\displaystyle
\left.\left.
\times
\varepsilon_{e}\Big(\mathbf{k}_{i}-\left(m_{1}-q_{0}\right)
\mathbf{k}_{0z}, \Omega\left(\mathbf{k}_{i}+q_{0}\mathbf{k}_{0z}\right)-m_{1}\omega_{0}\Big)\right)
\right]. 
\nonumber
\end{eqnarray}

The results of the numerical solution of the dispersion equation~(\ref{35}) for the reduced three-mode system~(\ref{33}), 
presented in Figures~\ref{fig1}--\ref{fig4}, show that for all unstable IC harmonics with $n=21,\,23,\,25,$ and $27$, driven by the HHFW with frequency $\omega_{0}=30.5\,\omega_{ci}$, 
the growth rate $\gamma=\mathrm{Im}\,\omega\left(\mathbf{k}_{i}\right)$ of the HHIC instability is of the same order as the frequency shift
$\delta\omega\left(\mathbf{k}_{i}\right)=\mathrm{Re}\,\omega\left(\mathbf{k}_{i}\right)-n\omega_{ci}$.
The nonlinear evolution of the detected HHIC instability can be 
investigated within the framework of the well-known renormalized theory of plasma turbulence~\cite{Dum}, 
which accounts for the nonlinear effect of ion scattering by an ensemble of unstable HHIC waves.
It follows from the solution~(\ref{38}) for the renormalized frequency $\hat{\omega}$, determined by Eq.~(\ref{27}),
\begin{eqnarray}
& \displaystyle
 \hat{\omega}=Re\, \hat{\omega} + i\gamma _{Nl}=\Omega\left(\mathbf{k}_{i}+q_{0}\mathbf{k}_{0z}\right)-q_{0}\omega_{0},
\label{40}
\end{eqnarray}
where $\gamma _{Nl}=Im\,  \hat{\omega}$, that instability is absent when $\gamma_{Nl}=0$ for the frequency 
$\Omega\left(\mathbf{k}_{i}\right)$ with $Im \Omega\left(\mathbf{k}_{i}\right)=0$. It follows from relation (\ref{27}) 
for the linear frequency $\omega$,
\begin{eqnarray}
& \displaystyle
Re\, \hat{\omega} + i\gamma _{Nl}=Re\, \omega+i\gamma_{Lin}-iC_{i},
\label{41}
\end{eqnarray}
that $\gamma_{Nl}=\gamma_{Lin}-C_{i}$, and the condition $\gamma_{Nl}=0$ 
corresponds to the balance $\gamma_{Lin}=C_{i}$ between the linear growth rate and the nonlinear damping resulting from 
ion scattering by IC parametric turbulence. For the potential 
$\varphi_{i}\left(\mathbf{k}_{i}+q_{0}\mathbf{k}_{0z},\hat{\omega}+q_{0}\omega_{0}\right)$ with solution~(\ref{38}),
\begin{eqnarray}
& \displaystyle C_{i}\left(\mathbf{k}_{i}+q_{0}\mathbf{k}_{0z}\right)=\frac{e_{i}^{2}}{m_{i}^{2}\omega_{ci}^{2}}
Re \sum\limits_{p_{1}=-\infty}^{\infty}\int d\mathbf{k}_{i1}|\varphi_{i}\left(\mathbf{k}_{i1}
+q_{0}\mathbf{k}_{0z}\right)|^{2}\mathcal{F}_{ip_{1}}\left( k_{i\bot}, k_{i1\bot}\right)
\nonumber
\\
&\displaystyle 
\times
\sqrt{\frac{\pi}{2}}\frac{1}{\left(k_{i1z}+q_{0}k_{0z}\right)v_{Ti}}
W\left(z_{ip_{1}}\left(\mathbf{k}_{i1}+q_{0}\mathbf{k}_{0z}\right)\right),
\label{42}
\end{eqnarray}
where  
\begin{eqnarray}
& \displaystyle
z_{ip_{1}}\left(\mathbf{k}_{i1}+q_{0}\mathbf{k}_{0z}\right)=\frac{\Omega\left(\mathbf{k}_{i1}
+q_{0}\mathbf{k}_{0z}\right)+iC_{i}\left(\mathbf{k}_{i}+q_{0}\mathbf{k}_{0z}\right)-q_{0}\omega_{0}
-p_{1}\omega_{ci}}{\sqrt{2}\left(k_{i1z}+q_{0}k_{z1}\right)v_{Ti}},
\label{43}
\end{eqnarray}
and $\mathcal{F}_{ip_{1}}\left(k_{i\bot},k_{i1\bot}\right)$ is determined by Eq.~(\ref{32}).

Equation~(\ref{43}) shows that the resonant exchange of energy between ions and HHIC waves 
continues until the ion--wave coherence is destroyed by nonlinear changes in the ion orbits 
resulting from the interaction of ions with the electric field of HHIC turbulence. 
In Eq.~(\ref{41}), this effect is represented by the term
$-iC_{i}\left(\mathbf{k}_{i}+q_{0}\mathbf{k}_{0z}\right)$.
In turbulent plasma, particles that are initially resonant may be driven out of resonance, whereas initially 
nonresonant particles may become resonant for a finite correlation time. When nonresonant ions become involved 
in the wave--particle interaction, the wave loses energy on average and is stabilized.
Analytically, this nonlinear resonance-broadening effect manifests itself through the growth of the linear
$|z_{p_{1}}\left(\mathbf{k}_{i1}+q_{0}\mathbf{k}_{0z}\right)|$ to values  larger than unity for the nonlinear
$|z_{ip_{1}}\left(\mathbf{k}_{i1}+q_{0}\mathbf{k}_{0z}\right)|$ determined by Eq. (\ref{43}).
This growth ceases when a steady-state level of the electrostatic potential
$\varphi_{in}\left(\mathbf{k}_{i}\right)$ of HHIC turbulence is reached. This level can be determined from Eq.~(\ref{42}), which, for
$\gamma_{NL}\left(\mathbf{k}_{i}+q_{0}\mathbf{k}_{0z}\right)=0$ and
$|z_{ip_{1}}\left(\mathbf{k}_{i1}+q_{0}\mathbf{k}_{0z}\right)|\gg1$, reduces to
\begin{eqnarray}
& \displaystyle 
C_{i}\left(\mathbf{k}_{i}+q_{0}\mathbf{k}_{0z}\right)=\frac{e_{i}^{2}}{m_{i}^{2}\omega_{ci}^{2}}
 \sum\limits_{p_{1}=-\infty}^{\infty}\int d\mathbf{k}_{i1}|\varphi_{i}\left(\mathbf{k}_{i1}
+q_{0}\mathbf{k}_{0z1}\right)|^{2}\mathcal{F}_{ip_{1}}\left( k_{i\bot}, k_{i1\bot}\right)
\nonumber
\\
&\displaystyle 
\times\left[\frac{C_{i}\left(\mathbf{k}_{i1}+q_{0}\mathbf{k}_{0z}\right)}{\left(\delta\Omega\left(\mathbf{k}_{i1}
+q_{0}\mathbf{k}_{01z}\right)\right)^{2}}+\sqrt{\frac{\pi}{2}}\frac{1}{\left(k_{i1z}+q_{0}k_{0z}\right)v_{Ti}}
e^{-z_{ip1}^{2}}\right],
\label{44}
\end{eqnarray}
where 
\begin{eqnarray}
& \displaystyle 
\left(\delta\Omega\left(\mathbf{k}_{i1}+q_{0}\mathbf{k}_{01z}\right)\right)^{2}=\left(\textit {Re}\,
\Omega\left(\mathbf{k}_{i1}+q_{0}\mathbf{k}_{01z}\right)-q_{0}\omega_{0}-p_{1}\omega_{ci}\right)^{2}.
\label{45}
\end{eqnarray}
The second term in Eq.~(\ref{44}) arises from resonant (quasilinear) diffusion, whereas the first term
 represents the resonance-broadening effect and is usually an order of magnitude larger.
Equation~(\ref{44}) is an integral equation for the spectral intensity
$|\varphi_{in}\left(\mathbf{k}_{i}\right)|^{2}$
of the potential corresponding to the maximum growth rate
$\gamma_{n}\left(\mathbf{k}_{i}\right)$.
The simplest order-of-magnitude estimate of the steady-state level of HHIC turbulence may be obtained
 by approximately replacing Eq.~(\ref{44}) by the algebraic equation for the root-mean-square spectral intensity
\begin{eqnarray}
&\displaystyle 
\bar{\varphi}_{i}=\left(\int |\varphi_{i}\left(\mathbf{k}_{i1}\right)|^{2}d\mathbf{k}_{i1}\right)^{1/2}.
\label{46}
\end{eqnarray} 
Employing the mean-value theorem for the integral over $\mathbf{k}_{i1}$ in Eq.~(\ref{44}) and assuming 
that the spectral intensity $|\varphi_{i}\left(\mathbf{k}_{i}\right)|^{2}$
is peaked at the values $\hat{k}_{ix}\rho_{i}$ and $\hat{k}_{iy}\rho_{i}$
corresponding to the most unstable linear wave number $\hat{\mathbf{k}}_{i}$,
the simplest order-of-magnitude estimate for the level of HHIC turbulence,
$e|\bar{\varphi}_{i}|/T_{i}$,
\begin{eqnarray}
&\displaystyle 
\frac{e_{i}|\bar{\varphi}_{i}|}{T_{i}}\sim 
\frac{1}{\omega_{ci}}\frac{|\delta \Omega |}{\rho^{2}_{i}\mathcal{F}_{i}^{1/2}
\left(\hat{k}_{i\bot }, \hat{k}_{i\bot }\right)}\sim \frac{1}{\hat{k}^{2}_{i\bot }\rho_{i}^{2}},
\label{47}
\end{eqnarray} 
follows from Eq.~(\ref{44}), where only the dominant terms with $p_{1}=1$ have been retained
 in the summations over $p_{1}$ in $\delta\Omega$ and $\mathcal{F}_{i}$.

\section{Heating of near-SOL plasma by parametric HHIC turbulence}\label{sec5}

The development of the ion-kinetic HHIC-quasimode decay instability identified in the present study leads to anomalous ion heating. 
The anomalous ion-heating rate is derived from the moments of the renormalized ion quasilinear equation, which accounts 
for the effect of ion scattering by HHIC turbulence,
\begin{eqnarray}
&\displaystyle \frac{\partial F_{i0}}{\partial t}
+\frac{e_{i}}{m_{i}\omega_{ci}}\left\langle \frac{\partial\varphi_{i}}{\partial \bar{X}_{i}}
\frac{\partial f_{i}}{\partial \bar{Y}_{i}}-\frac{\partial\varphi_{i}}{\partial \bar{Y}_{i}}\frac{\partial f_{i}}{\partial 
\bar{X}_{i}}\right\rangle
+\frac{e_{i}}{m_{i}}\frac{\omega_{ci}}{\bar{v}_{i\bot}}\left\langle \frac{\partial\varphi_{i}}{\partial \bar{\phi}}
\frac{\partial f_{i}}{\partial \bar{v}_{i\bot}}-\frac{\partial\varphi_{i}}{\partial \bar{v}_{i\bot}}\frac{\partial f_{i}}{\partial 
\bar{\phi}}\right\rangle
\nonumber
\\ 
&\displaystyle
-\frac{e_{i}}{m_{i}}\left\langle\frac{\partial\varphi_{i}}{\partial z_{i}}\frac{\partial f_{i}}{\partial \bar{v}_{iz}}
\right\rangle=0.
\label{48}
\end{eqnarray}
In Eq. (\ref{48}),  $f_{i}\left(\bar{v}_{i\bot}, \bar{\phi}, \bar{X}_{i},  \bar{Y}_{i}, z, t\right)$ and potential  
$\varphi_{i}\left(\bar{v}_{i\bot}, \bar{\phi}, \bar{X}_{i},  \bar{Y}_{i}, z, t\right)$ are determined by the equations
\begin{eqnarray}
&\displaystyle
f_{i}\left(\bar{v}_{i\bot}, \bar{\phi}, \bar{X}_{i},  \bar{Y}_{i}, z, t\right)=i\frac{e_{i}}{m_{i}}
\int \limits_{t_{0}}^{t}dt_{1}\int d\mathbf{k}_{i}d\omega
\sum_{p=-\infty}^{\infty}\left[\frac{k_{iy}}{\omega_{ci}}\frac{\partial F_{i}}{\partial \bar{X}_{i}}
+\frac{p\omega_{ci}}{\bar{v}_{i\bot}}\frac{\partial F_{i}}{\partial \bar{v}_{i\bot}}+k_{iz}\frac{\partial F_{i}}{\partial 
\bar{v}_{iz}}\right]
\nonumber
\\ 
&\displaystyle
\times\varphi_{i}\left(\mathbf{k}_{i}, \omega\right)J_{p}\left(\frac{k_{i\bot}\bar{v}_{i\bot}}{\omega_{ci}}\right)
\exp\left(-i\omega t_{1}+ik_{ix}\bar{X}_{i}+ik_{iy}\bar{Y}_{i}-ik_{iz}\bar{v}_{iz}\left(t-t_{1}\right)\right.
\nonumber
\\ 
&\displaystyle
\left.
-ip\left(\bar{\phi}-\omega_{ci}t_{1}\right)-C_{i}\left(\mathbf{k}_{i}\right)t_{1}\right),
\label{49}
\\
&\displaystyle
\varphi_{i}\left(\bar{v}_{i\bot}, \bar{\phi}, \bar{X}_{i},  \bar{Y}_{i}, z, t\right)=\sum_{p_{1}=-\infty}^{\infty}\int 
d\mathbf{k}_{i1}\,d\omega\,\varphi_{i}\left(\mathbf{k}_{i1}, 
\omega\right)J_{p_{1}}\left(\frac{k_{i1\bot}\bar{v}_{i\bot}}{\omega_{ci}}\right)
\nonumber
\\ 
&\displaystyle
\times \exp\left(-i\omega t+ik_{ix1}\bar{X}_{i}+ik_{iy1}\bar{Y}_{i}-ik_{1z}z_{i}
-ip_{1}\left(\bar{\phi}-\omega_{ci}t_{1}\right)-C_{i}\left(\mathbf{k}_{i1}\right)t\right).
\label{50}
\end{eqnarray}
In the case where HHIC turbulence is powered by the IC-quasimode decay instability, the potential
$\varphi_{i}\left(\mathbf{k}_{i},\omega\right)$ in Eqs.~(\ref{49}) and~(\ref{50}) is coupled to the quasimode potential
\begin{eqnarray}
&\displaystyle
\varphi_{i}\left(\mathbf{k}_{i}+q_{0}\mathbf{k}_{0z}, 
\omega+q_{0}\omega_{0}\right)
\nonumber
\\ 
&\displaystyle
=\varphi_{i}\left(\mathbf{k}_{i}+q_{0}\mathbf{k}_{0z}\right)
\delta\left(\omega+q_{0}\omega_{0}-\Omega\left(\mathbf{k}_{i}+q_{0}\mathbf{k}_{0z}\right)
-iC\left(\mathbf{k}_{i}+q_{0}\mathbf{k}_{0z}\right)\right)
\label{51}
\end{eqnarray}
by Eq. (\ref{25})
\begin{eqnarray}
&\displaystyle
\varphi_{i}\left(\mathbf{k}_{i}, \omega\right)= -\frac{1}{\varepsilon \left(\mathbf{k}_{i}, \omega\right)}
\sum\limits_{m=-\infty}^{\infty}J_{m}\left(a_{ei}\right)J_{m+q}\left(a_{ei}\right)
\varepsilon_{e} \left(\mathbf{k}_{i}-m\mathbf{k}_{iz}, \omega-m\omega_{0}\right)e^{iq_{0}\delta}
\nonumber
\\ 
&\displaystyle
\times\varphi_{i}\left(\mathbf{k}_{i}+q_{0}\mathbf{k}_{0z}\right)
\delta\left(\omega+q_{0}\omega_{0}-\Omega\left(\mathbf{k}_{i}+q_{0}\mathbf{k}_{0z}\right)
-iC\left(\mathbf{k}_{i}+q_{0}\mathbf{k}_{0z}\right)\right),
\label{52}
\end{eqnarray}
for which Eq.~(\ref{37}) holds. In this case, only the term with $q=q_{0}$ is retained in the summation over $q$. 
With the quasimode potential~(\ref{51}), the renormalized quasilinear equation~(\ref{48}) becomes
\begin{eqnarray}
&\displaystyle
 \frac{\partial F_{i0}}{\partial t}=\frac{\pi e_{i}^{2}}{m^{2}_{i}}\int 
d\mathbf{k}_{i}\sum\limits_{m_{1}=-\infty}^{\infty}\sum\limits_{m_{2}=-
\infty}^{\infty}\sum\limits_{p=-\infty}^{\infty}J_{m_{1}}\left(a_{ei}\right)J_{m_{1}+q_{0}}\left(a_{ei}\right)
J_{m_{2}}\left(a_{ei}\right)J_{m_{2}-q_{0}}\left(a_{ei}\right)
\nonumber
\\ 
&\displaystyle
\times|\varphi_{i}\left(\mathbf{k}_{i}+q_{0}\mathbf{k}_{0z}\right)|^{2}\frac{\varepsilon_{e}\left(\mathbf{k}_{i}
-m_{1}\mathbf{k}_{0z},  \omega\left(\mathbf{k}_{i}+q_{0}\mathbf{k}_{0z}\right)
-\left(m_{1}+q_{0}\right)\omega_{0}\right)
}{|\varepsilon\left(\mathbf{k}_{i}, \omega\left(\mathbf{k_{i}}+q_{0}\mathbf{k}_{0z}\right)-q_{0}\omega_{0}\right)|^{2}}
\nonumber
\\ 
&\displaystyle
\times\varepsilon_{e}\left(-\mathbf{k}_{i}+m_{2}\mathbf{k}_{0z}, -\omega\left(\mathbf{k}_{i}+q_{0}\mathbf{k}_{0z}\right)
-\left(m_{1}-q_{0}\right)\omega_{0}\right)\nonumber
\\ 
&\displaystyle
\times\left(\frac{p\omega_{ci}}{\bar{v}_{i\bot}}\frac{\partial}{\partial\bar{ v}_{i\bot}}+\frac{k_{iy}}{\omega_{ci}}
\frac{\partial}{\partial \bar{X}_{i}}+k_{iz}\frac{\partial}{\partial \bar{v}_{z}}\right)J^{2}_{p}\left(\frac{k_{i\bot}
\bar{v}_{i\bot}}{\omega_{ci}}\right)R_{i}\left(\mathbf{k}_{i}+q_{0}\mathbf{k}_{0z},\bar{v}_{z}\right)
\nonumber
\\ 
&\displaystyle
\times \left(\frac{p\omega_{ci}}{\bar{v}_{i\bot}}\frac{\partial F_{i0}}{\partial 
\bar{v}_{i\bot}}+\frac{k_{iy}}{\omega_{ci}}\frac{\partial F_{i0}}{\partial \bar{X}_{i}}+k_{iz}\frac{\partial 	F_{i0}}{\partial 
\bar{v}_{z}}\right).
\label{53}
\end{eqnarray}
In Eq.~(\ref{53}), $R_{i}$ is the renormalized resonance function defined by
\begin{eqnarray}
&\displaystyle 
R_{i}\left(\mathbf{k}_{i}+q_{0}\mathbf{k}_{0z}, \bar{v}_{z}\right)=Re\,\sum_{p_{1}=-\infty}^{\infty} \int\limits^{\infty}_{0}d\tau 
e^{-i\left(\delta\Omega\left(\mathbf{k}_{i}+q_{0}\mathbf{k}_{0z}\right)-\left(k_{iz}+q_{0}k_{0z}\right)v_{iz}
+iC_{i}\left(\mathbf{k}_{i}+q_{0}\mathbf{k}_{0z}\right)\right)\tau}
\nonumber
\\ 
&\displaystyle
=R_{iNl}\left(\mathbf{k}_{i}+q_{0}\mathbf{k}_{0z}, \bar{v}_{z}\right)+R_{iQuasy}\left(\mathbf{k}_{i}+q_{0}\mathbf{k}_{0z}, 
\bar{v}_{z}\right)
\nonumber
\\ 
&\displaystyle
=\sum_{p_{1}=-\infty}^{\infty}\left(\frac{C_{i}\left(\mathbf{k}_{i}+q_{0}\mathbf{k}_{0z}\right)}
{\left(\delta\Omega\left(\mathbf{k}_{i}+q_{0}\mathbf{k}_{0z}\right)
-\left(k_{iz}+q_{0}k_{0z}\right)v_{iz}\right)^{2}+C^{2}_{i}\left(\mathbf{k}_{i}+q_{0}\mathbf{k}_{0z}\right)}\right.
\nonumber
\\ 
&\displaystyle
-i\pi\delta\left(\delta\Omega\left(\mathbf{k}_{i}
+q_{0}\mathbf{k}_{0z}\right)
-\left(k_{iz}+q_{0}k_{0z}\right)v_{iz}\right)\Big ) ,
\label{54}
\end{eqnarray}
where $\tau=t-t_{1}$, and the time interval $\tau>\gamma\left(\mathbf{k}_{i}\right)^{-1}$ is much longer than the correlation time. 
The parameter $z_{in}$ is determined by Eq.~(\ref{43}).
The function $R_{i}$ represents the renormalized counterpart of the resonant denominator 
$\left(\omega\left(\mathbf{k}_{i}\right)-n\omega_{ci}-k_{z}v_{iz}\right)^{-1}$,
which appears in the conventional quasilinear theory.

The temporal and spatial evolution of  the  average perpendicular energy density of ions,  
\begin{eqnarray*}
&\displaystyle n_{i0}T_{i\bot}\left(X_{i}, t\right)=\frac{1}{2}\int d\mathbf{v}_{i}F_{i0}\left(\mathbf{v}_{i}, X_{i}, t\right)
m_{i}v^{2}_{i\bot},
\end{eqnarray*}
is resulted from the interactions of ions with HHIC turbulence in the inhomogeneous plasmas, effect of which is determined 
by the first term in Eq. (\ref{54}).  The corresponding equation is derived by multiplying 
Eq. (\ref{53}) on $m_{i}v^{2}_{i\bot}/2$ and integrated over velocities. Integrating by parts over $v_{i\bot}$, and 
employing the relation in Eq. (\ref{53})
\begin{eqnarray}
&\displaystyle
Im \,\varepsilon_{i}\left(\mathbf{k}_{i}, \Omega\left(\mathbf{k}_{i}
+q_{0}\mathbf{k}_{0z}\right)-q\omega_{0}\right)=-\frac{4\pi^{2}e_{i}^{2}}{m_{i}k^{2}}\int d\mathbf{v}_{i}
\sum_{p=-\infty}^{\infty}J^{2}_{p}\left(\frac{k_{i\bot}v_{i\bot}}{\omega_{ci}}\right)
\nonumber
\\ 
&\displaystyle
\times\left(\frac{k_{iy}}{\omega_{ci}}\frac{\partial F_{i0}}{\partial \bar{X}_{i}}+\frac{p\omega_{ci}}{v_{i\bot}}\frac{\partial 
F_{i0}}{\partial v_{i\bot}}\right)R_{iNl}\left(\mathbf{k}_{i}+q_{0}\mathbf{k}_{0z}, \bar{v}_{z}\right),
\label{55}
\end{eqnarray}
we derive the equation
\begin{eqnarray}
&\displaystyle \frac{\partial }{\partial t}n_{0i}T_{i\bot}\approx {\nu_{i\bot}} n_{0i}T_{i\bot}
+\chi_{i\bot}\frac{\partial^{2}}{\partial X_{i}^{2}} n_{0i}T_{i\bot},
\label{56}
\end{eqnarray}
in which
\begin{eqnarray}
&\displaystyle 
\nu_{i\bot}\sim\gamma\left(\hat{\mathbf{k}}_{i}\right)\frac{\mathcal{W}}{n_{0i}T_{i\bot}},
\label{57}
\end{eqnarray}
is the ion turbulent heating rate across the magnetic field,
\begin{eqnarray}
&\displaystyle 
\chi_{i\bot}\sim\frac{\gamma\left(\hat{\mathbf{k}}_{i}\right)}{\hat{k}^{2}_{i\bot}}\left(\frac{\omega_{ci}}
{\delta \Omega\left(\hat{\mathbf{k}}_{i}\right)}\right)^{2}\frac{\mathcal{W}}{n_{0i}T_{i\bot}},
\label{58}
\end{eqnarray} 
is coefficient of the anomalous ion temperature conductivity across the magnetic field.  ${\mathcal{W}}$ is  the energy 
density of  HHIC turbulence in the saturated state, which is determined by the relation
\begin{eqnarray}
&\displaystyle
\mathcal{W}=\int \mathcal{W}\left(\mathbf{k}_{i}\right)d\mathbf{k}_{i}
\approx \hat{k}^{2}_{i}\omega\left(\hat{\mathbf{k}}_{i}\right)
\frac{1}{4\pi}\left|\bar{\varphi}_{i}\right|^{2}\frac{\partial \varepsilon_{i}}
{\partial \omega\left(\hat{\mathbf{k}}_{i}\right)},	
\label{59}
\end{eqnarray}
where the root-mean-square magnitude $\left|\bar{\varphi}_{i}\right|$ of the IC turbulence 
potential in the saturated state is estimated by Eq. (\ref{47}).  

The ion heating along the magnetic field is resulted from the  interactions of ions with HHIC turbulence
 under the condition of ion-cyclotron resonance. Under this condition, the ion resonance function~(\ref{54}) is determined 
 primarily by the term $R_{iQuasy}\left(\mathbf{k}_{i}+q_{0}\mathbf{k}_{0z},\bar{v}_{z}\right)$.
The estimate for the ion-heating rate along the magnetic field,
\begin{eqnarray}
&\displaystyle 
n_{0}\frac{\partial T_{iz}}{\partial t}\sim k_{z}\rho_{i}\gamma\left(\hat{\mathbf{k}}_{i}\right) 
\frac{\mathcal{W}}{n_{0i}T_{iz}}T_{iz},
\label{60}
\end{eqnarray} 
 is derived by multiplying 
Eq.~(\ref{53}) by $m_{i}v_{z}^{2}/2$ and integrating over the ion velocities $\mathbf{v}_{i}$.
It follows from Eqs.~(\ref{57}) and~(\ref{60}) that the absorption of HHIC parametric turbulence 
by ions in near-SOL plasma results in anisotropic ion heating, with
\begin{eqnarray}
&\displaystyle 
\left. \frac{\partial T_{iz }}{\partial t}\right/\frac{\partial T_{i\bot }}{\partial t}\sim k_{z}\rho_{i}\ll 1.
\label{61}
\end{eqnarray}

\section{Conclusions}\label{sec6}

In this paper, we have developed a theory of electrostatic parametric instabilities driven by high-power HHFW in the inhomogeneous near-SOL 
tokamak plasma. The model extends previous studies of uniform SOL plasmas by incorporating finite density and temperature gradients. 
Numerical analysis of the reduced three-mode dispersion equation (\ref{35}) demonstrates the existence of unstable high-harmonic 
ion-cyclotron (HHIC) modes driven by the parametric decay of HHFW with frequency $\omega_{0}=30.5\,\omega_{ci}$ into an HHIC 
(Bernstein) wave and an HHIC quasimode.

It was found that the HHIC-quasimode decay instability is excited only within a limited interval of ion-cyclotron harmonic numbers. For 
$\omega_{0}=30.5\,\omega_{ci}$, unstable solutions exist only for the high-harmonic modes with $n=17$--$27$, while the largest 
growth rates are obtained for the intermediate harmonics. The instability is most efficiently excited in plasmas with steep density gradients. 
Ion-temperature-gradient effects provide the dominant drive for the instability, whereas electron-temperature-gradient effects remain 
comparatively weak. The unstable modes develop in a regime where inverse ion Landau damping is effective and constitutes the principal 
mechanism responsible for the instability growth.

A renormalized theory of the ion-kinetic HHIC-quasimode decay instability, which accounts for  the average nonlinear effect of ion 
scattering by the electrostatic turbulent field,  has been developed  to describe the nonlinear evolution 
and saturation of the instability.  The saturation level of HHIC turbulence is determined by the balance between the linear 
growth and nonlinear damping associated with ion scattering by the HHIC turbulent electric field.

The developed HHIC turbulence produces strongly anisotropic ion heating in the near-SOL plasma, with the perpendicular ion heating rate 
greatly exceeding the parallel ion heating rate. This result is consistent with NSTX experimental observations \cite{Taylor,Biewer,Hosea}. 
The obtained results indicate that HHIC parametric turbulence provides an efficient nonlinear channel for RF-power 
absorption and edge-ion heating in the near-SOL region during HHFW heating and current-drive experiments.

Because both the growth rate of the instability and the anomalous ion heating rate in the near-SOL plasma are approximately an order 
of magnitude larger than those in the SOL plasma, the HHIC parametric turbulence in the near-SOL plasma is expected to play an increasingly 
important role in RF-power absorption and edge ion heating in plasmas with a well-developed pedestal. These results suggest that pedestal 
formation, while beneficial for plasma confinement, may enhance nonlinear RF-power absorption and anomalous ion heating in the tokamak 
edge region, thereby complicating the achievement of the non-inductively sustained steady-state tokamak operation.
 
\begin{acknowledgments}
This work was supported by the National Research Foundation of Korea (NRF) through the National R\&D 
Program funded by the Ministry of Education, Science and Technology under Grant No.~NRF-2022R1A2C1012808126 
\end{acknowledgments}

\bigskip
{\bf DATA AVAILABILITY}

\bigskip
The data that support the findings of this study are available from the corresponding 
author upon reasonable request.

\end{document}